\documentclass{aa}

\usepackage{amsmath}
\usepackage{amssymb}
\usepackage[usenames, dvipsnames]{color}
\usepackage{footmisc}
\usepackage[modulo,switch]{lineno}
\usepackage{graphicx}
\usepackage{multirow}
\usepackage{natbib}
\usepackage{nicefrac}
\usepackage{pdfpages}
\usepackage[rightcaption]{sidecap}
\usepackage{units}
\usepackage{url}
\usepackage[normalem]{ulem}
\usepackage{xcolor}
\usepackage{txfonts}
\usepackage{hyperref}

\hypersetup{
    colorlinks=true,
    linkcolor=blue,
    citecolor=blue,
    urlcolor=blue,
    pdftitle={Solar H-alpha excess during Solar Cycle 24 from full-disk filtergrams of the Chromospheric Telescope},
    pdfauthor={Diercke et al.}
}

\makeatletter
\renewcommand*\aa@pageof{, page \thepage{} of \pageref*{LastPage}}
\makeatother

\newcommand\arcdeg{\mbox{$^\circ$}}

\begin{document} 

   \title{Solar H$\alpha$ excess during Solar Cycle 24 from full-disk filtergrams of the Chromospheric Telescope}
   
   \titlerunning{Solar H$\alpha$ excess during Solar Cycle 24}
   
   \subtitle{}

   \author{A. Diercke\inst{1, 2,3}
          \and
          C. Kuckein\inst{1,4,5}
          \and
          P. W. Cauley\inst{6}
          \and
          K. Poppenh\"ager\inst{1}
          \and
           J. D. Alvarado-G\'omez\inst{1}
           \and
           E. Dineva\inst{1,2}
          \and 
          C. Denker\inst{1}
          }

   \institute{Leibniz-Institut f\"ur Astrophysik Potsdam (AIP),
              An der Sternwarte 16,
              14482 Potsdam, Germany
        \and
             Universit\"at Potsdam,
             Institut f\"ur Physik und Astronomie,
             Karl-Liebknecht-Stra\ss{}e 24/25,
             14476 Potsdam, Germany
        \and 
            National Solar Observatory (NSO), 
            3665 Discovery Drive,
            Boulder, CO, USA, 80303\\
            \email{adiercke@nso.edu}
        \and
            Instituto de Astrof\'{i}sica de Canarias (IAC), 
            V\'{i}a L\'{a}ctea s/n, 38205 La Laguna, Tenerife, Spain
        \and 
            Departamento de Astrof\'{\i}sica, Universidad de La Laguna
            38205, La Laguna, Tenerife, Spain 
        \and
            Laboratory for Atmospheric and Space Physics,
            University of Colorado Boulder, 
            Boulder, CO 80303
        }

   \date{Received December 08, 2020; accepted February 17, 2022}

% \abstract{}{}{}{}{} 
% 5 {} token are mandatory
 
  \abstract
  % context heading (optional)
  % {} leave it empty if necessary  
   {The chromospheric H$\alpha$ spectral line  is a strong line in the spectrum of the Sun and other stars. In the stellar regime, this spectral line is already used as a powerful tracer of stellar activity. For the Sun,  other tracers, such as \mbox{Ca\,\textsc{ii}}~K, are typically used to monitor solar activity. Nonetheless, the Sun is observed constantly in H$\alpha$ with globally distributed ground-based full-disk imagers.}
  % aims heading (mandatory)
   {The aim of this study is to introduce the imaging H$\alpha$ excess and deficit as tracers of solar activity and compare them to other established indicators. Furthermore, we investigate whether the active region coverage fraction or the changing H$\alpha$ excess in the active regions dominates temporal variability in solar H$\alpha$ observations.
   }
  % methods heading (mandatory)
   {We used observations of full-disk H$\alpha$ filtergrams of the Chromospheric Telescope (ChroTel) and morphological image processing techniques to extract the imaging H$\alpha$ excess and deficit, which were derived from the intensities above or below 10\% of the median intensity in the filtergrams, respectively. These thresholds allowed us to filter for bright features (plage regions) and dark absorption features (filaments and sunspots). In addition, the thresholds were used to calculate the mean intensity $I^\mathrm{E/D}_\mathrm{mean}$ for H$\alpha$ excess and deficit regions. We describe the evolution of the H$\alpha$ excess and deficit during Solar Cycle~24 and compare it to the mean intensity and other well established tracers: the relative sunspot number, the F10.7\,cm radio flux, and the \mbox{Mg\,\textsc{ii}} index. In particular, we tried to determine how constant the H$\alpha$ excess and number density of H$\alpha$ excess regions are between solar maximum and minimum. The number of pixels above or below the intensity thresholds were used to calculate the area coverage fraction of H$\alpha$ excess and deficit regions on the Sun, which was compared to the imaging H$\alpha$ excess and deficit and the respective mean intensities averaged for the length of one Carrington rotation. In addition, we present the H$\alpha$ excess and mean intensity variation of selected active regions during their disk passage in comparison to the number of pixels of H$\alpha$ excess regions.
   }
  % results heading (mandatory)
   {The H$\alpha$ excess and deficit follow the behavior of the solar activity over the course of the cycle. They both peak around solar maximum, whereby the peak of the H$\alpha$ deficit is shortly after the solar maximum. Nonetheless, the correlation of the monthly averages of the H$\alpha$ excess and deficit is high with a Spearman correlation of $\rho = 0.91$. The H$\alpha$ excess is closely correlated to the chromospheric \mbox{Mg\,\textsc{ii}} index with a correlation of 0.95. The highest correlation of the H$\alpha$ deficit is found with the F10.7\,cm radio flux, with a correlation of 0.89, due to their peaks after the solar activity maximum. Furthermore, the H$\alpha$ deficit reflects the cyclic behavior of polar crown filaments and their disappearance shortly before the solar maximum. We investigated the mean intensity distribution for H$\alpha$ excess regions for solar minimum and maximum. The shape of the distributions for solar minimum and maximum is very similar, but with different amplitudes. Furthermore, we found that the area coverage fraction of H$\alpha$ excess regions and the H$\alpha$ excess are strongly correlated with an overall Spearman correlation of 0.92. The correlation between the H$\alpha$ excess and the mean intensity of H$\alpha$ excess regions is 0.75. The correlation of the area coverage fraction and the mean intensity of H$\alpha$ excess regions is in general relatively low ($\rho = 0.45$) and only for few active regions is this correlation above 0.7. The weak correlation between the area coverage fraction and mean intensity leaves us pessimistic that the degeneracy between these two quantities can be broken for the modeling of unresolved stellar surfaces.
   }
  % conclusions heading (optional), leave it empty if necessary 
   {}
   
%maximum 6 key words
 
   \maketitle
%
%-------------------------------------------------------------------

\section{Introduction}

The first record of a varying sunspot number throughout a solar cycle was found in systematic sunspot observations by S.~H.\ Schwabe in the 19th century \citep{Schwabe1844, Hathaway2010, Arlt2013, Arlt2020}. This was followed in systematic observations by R.\ Carrington \citep{Carrington1858} and G.\ Sp\"orer \citep[e.g.,][]{Spoerer1879, Diercke2015}. R.\ Wolf established the relative sunspot number, counting the spot groups and individual sunspots \citep{Hathaway2010}. This method is still used nowadays to determine the international sunspot number provided by the Solar Influences Data Analysis Center (SIDC) of the Royal Observatory of Belgium, which is used to monitor the 11-year activity cycle of the Sun.

In addition to white-light observations of the Sun, the solar activity was monitored in different wavelengths, for example, the chromospheric \mbox{Ca\,\textsc{ii}\,K} line, whose emission in plage regions is correlated to the magnetic field with a power-law relation with an exponent of about 0.5 \citep{Schrijver1989, Rezaei2007, Barczynski2018}. \citet{Livingston2007} summarized the results from three solar cycles where spectroscopic data of different chromospheric lines were analyzed. The spectral \mbox{Ca\,\textsc{ii}\,K} index showed the highest variations throughout the cycle with amplitudes of about 25\%.  The higher intensity contrast is partly due to the steep Planck function and, therefore, stronger temperature response at short wavelengths around 4000\,\AA\ \citep{Ayres1989, Sheminova2012}. Other chromospheric line indicators, including the \mbox{He\,\textsc{i}}~10830\,\AA\ equivalent width,  \mbox{Ca\,\textsc{ii}}~8542\,\AA\ central depth, H$\alpha$ central depth, or CN bandhead at 3883\,\AA, all showed variations with the solar cycle, but with a lower relative amplitude as compared to \mbox{Ca\,\textsc{ii}\,K} \citep{Livingston2007, Ermolli2015}. Full-disk \mbox{Ca\,\textsc{ii}\,K} measurements show an average increase of 30\% of the central intensity between minimum and maximum of the solar cycle \citep{White1981}, whereby the increase is associated with solar plages. \citet{White1981} suggested that the variability of the quiet-Sun network is negligible in the rising phase of the solar cycle. A more recent study by \citet{Bertello2016} inspects the correlation of the Ca\,\textsc{ii}~K emission index and the sunspot number. They found a strong correlation on timescales of months or years, which they explain by the high correlation of both quantities to the magnetic flux. \citet{Naqvi2010} used average full-disk images of the Big Bear Solar Observatory \citep[BBSO,][]{Denker1999} to calculate an index using an adaptive intensity threshold to extract the bright plage regions from \mbox{Ca\,\textsc{ii}\,K} full-disk images. More details on the method can be found in \citet{Johannesson1998}. With this index, they characterized the solar activity of Solar Cycle~23 for different thresholds in comparison with other solar indices. A very complete data collection of \mbox{Ca\,\textsc{ii}}~K observations was recently presented in \citet{Chatzistergos2020a}, who collected around 290\,000 full-disk \mbox{Ca\,\textsc{ii}}~K observations from 43 data sets including observations between 1892 and 2019. They used the fractional plage area on the solar disk as an indicator for the solar activity.

Other well-established solar indices include \mbox{Mg\,\textsc{ii}} and the F10.7\,cm radio flux. The chromospheric \mbox{Mg\,\textsc{ii}} doublet at 2795.6\,\AA\ and 2802.7\,\AA\ has been used since 1978 as a proxy for solar activity \citep{Viereck1999}. The index is calculated from the core-to-wing ratio of the spectral line. This spectral \mbox{Mg\,\textsc{ii}} index varies up to 20\% between minimum and maximum. It is especially good in monitoring solar faculae, as \citet{Lean1997} determined from solar irradiance models, which consider both sunspot darkening and faculae brightening. The F10.7\,cm solar flux index has monitored the disk integrated radio emission at 10.7\,cm or 2800\,MHz since 1946 \citep{Tapping1994, Hathaway2010}. Several measurements a day are obtained, whereby periods of flares are avoided. The radio flux is strongly correlated with the relative sunspot number, but with a lag time of about one month \citep{Hathaway2010}.

Single active regions can influence the total solar irradiance. The lifetime of active regions can vary from a few hours for small ephemeral regions to months for large active regions \citep{vanDrielGesztelyi2015}. Furthermore, the active regions can evolve or decay during the disk passage. \citet{Toriumi2020} studied transiting active regions with space-borne data from the Helioseismic and Magnetic Imager \citep[HMI,][]{Scherrer2012} and the Atmospheric Imaging Assembly \citep[AIA,][]{Lemen2012} onboard the Solar Dynamics Observatory \citep[SDO,][]{Pesnell2012} as well as data of the Hinode X-Ray Telescope \citep[XRT,][]{Golub2007}. The continuum observations showed that the light curves of the active region's intensity decrease when it is at disk center, while the light curve increases when the active region is close to the limb. This effect is well known from observations of the total solar irradiance \citep[TSI,][]{Froehlich2012} and shows that the effects of the umbra and penumbra of the active region dominates close to disk center. The line-of-sight magnetic flux as well as the UV irradiance (1600\,\AA\ and 1700\,\AA) show a bell-shaped (or mountain-shaped) curve during the disk passage. Thereby, UV wavelengths are more sensitive to faculae regions rather than to sunspots \citep{Simoes2019, Toriumi2020}.

During the solar cycle, so-called active longitudes are established on the solar surface. These are longitudes, where sunspots predominantly form \citep{Berdyugina2003, Hathaway2010, Kramynin2021}. They can be studied by creating Carrington reference frames for each solar rotation.  \citet{Berdyugina2003} discovered that the active longitudes on the northern and southern hemisphere are separated by about 180\arcdeg. The location of the active longitude changes between solar cycles, but it can be a persistent structure for up to two decades \citep{Hathaway2010}.

The Sun is regularly observed in another prominent chromospheric line, that is to say the H$\alpha$ spectral line at $6562.8$\,\AA. Several observatories form the Global H$\alpha$ Network \citep{Steinegger2000}, including the BBSO and Kanzelh\"ohe Solar Observatory \citep[KSO,][]{Otruba1999},  which take regular full-disk images in the line core of H$\alpha$. In total, seven facilities around the world are part of this network, which continuously observe H$\alpha$ line-core filtergrams. Similarly, the Global Oscillation Network Group \citep[GONG,][]{Harvey1996} of the U.S.\ National Solar Observatory (NSO) consists of six facilities, which produce, among other data products, time series of full-disk H$\alpha$ filtergrams and photospheric magnetograms. In addition, other observatories provide regular full-disk H$\alpha$ observations, for example, the Solar Flare Telescope in Mitaka, Japan \citep{Hanaoka2020} and the Chromospheric Telescope in Tenerife, Spain \citep[ChroTel,][]{Kentischer2008, Bethge2011}. The large number of ground-based facilities all around the world facilitate almost continuous coverage of the Sun with full-disk H$\alpha$ filtergrams, bridging bad weather locally and the day-night cycle.

The variation of H$\alpha$ throughout the solar cycle is known from disk-integrated observations \citep{Livingston2007, Maldonado2019}. The H$\alpha$ line has the advantage of displaying filaments alongside sunspots as dark absorption features as well as plage regions around active regions as bright features. This is different to other chromospheric activity indicators such as \mbox{He\,\textsc{i}}\,10830\,\AA, where plage and filaments are both seen in absorption. The influence of filaments on the integrated intensity of H$\alpha$ observations was discussed in \citet{Meunier2009}. They found a higher correlation between the \mbox{Ca\,\textsc{ii}}\,K index and the H$\alpha$ integrated flux during the ascending phase of the cycle compared to the minimum, which they explained is due to the influence of absorption features on the solar surface.

Filaments are the dominating absorption features on the solar disk in H$\alpha$. They appear in all sizes, reaching from small-scale active region filaments to large-scale quiet Sun filaments, which can reach sizes of more then half a solar diameter \citep[e.g.,][]{Kuckein2016, Diercke2018}. A special type of quiet-Sun filaments are polar crown filaments, which are located at high latitudes \citep{Leroy1983, Leroy1984}. They form at the interface of unipolar flux at the poles and flux of opposite polarity, which is transported from active regions close to the equator toward the pole. This polarity inversion line (PIL) appears at mid-latitudes shortly after the maximum of the solar cycle and it appears close to the pole shortly before solar maximum, which is called "rush-to-the-pole" \citep{Cliver2014, Xu2018, Diercke2019b, Xu2021}. Polar crown filaments disappear from the solar disk at around solar maximum, when the magnetic field reversal happens.  To better understand the influence of absorption features on the solar integrated intensity, an analysis of the cyclic behavior of absorption features during the solar cycle is important.

Another advantage of using H$\alpha$ as a solar activity indicator is that H$\alpha$ spectra are easily obtainable for stars other than the Sun. Thus, activity metrics and cycle information based on H$\alpha$ variations on the Sun can potentially be used to infer stellar activity properties for which \mbox{Ca\,\textsc{ii}\,H\,\&\,K} diagnostics, for example, are not obtainable.

The Sun is the only star for which we can obtain spectra of active regions at high spatial resolution. As a result, there is an inherent degeneracy for other stars between the projected area and intrinsic emission strength of their active regions \citep[e.g.,][]{cauley17}. This degeneracy makes it difficult to measure such properties for other stars, which is important when trying to quantify and remove the contamination of stellar activity from exoplanet transmission spectra \citep{cauley18, rackham18}. It is also important from the view of stellar physics to infer the intrinsic emission strengths in individual active regions for different chromospheric lines for stars of varying spectral type and magnetic activity level. Empirical constraints on these quantities will provide guidance for sophisticated 3D magnetohydrodynamic (MHD) radiative transfer models of stellar atmospheres \citep{Freytag2002, leenaarts12, Holzreuter15, stepan15, Haberreiter2021}.

The aim of this study is to explore the usefulness of the spatially resolved H$\alpha$ excess as a tracer of solar activity. Bright and dark regions, each at different thresholds, are identified in ChroTel full-disk H$\alpha$ filtergrams. We compare the imaging H$\alpha$ excess and deficit with established solar activity tracers such as the relative sunspot number, the F10.7\,cm radio flux, and the \mbox{Mg\,\textsc{ii}} index (Sect.~\ref{sec:halpha_index}). One goal is to determine the contribution of H$\alpha$ excess regions  during low solar activity to solar cycle variations of the H$\alpha$ spectral irradiance and to relate the distribution to some of the assumptions that are made in X-ray studies of stellar activity cycles. (Sect.~\ref{sec:int_dist}).  In addition, we address the question of whether the H$\alpha$ activity level is dominated either by changes in the mean intensity of active regions or by the area coverage fraction of H$\alpha$ excess regions on the solar disk (Sect.~\ref{sec:index_rotation} and Sect.~\ref{sec:index_ar}). This is especially relevant for attempting to disentangle the area coverage fraction and active region emission strength for unresolved stellar surfaces.

%--------------------------------------------------------------------
\section{Observations and data processing} \label{sec:obs}

The Chromospheric Telescope \citep[ChroTel, ][]{Kentischer2008, Bethge2011} is a robotic ground-based 10-centimeter telescope, located at the Observatorio del Teide in Tenerife, Spain. It was built and is operated by the Leibniz Institute for Solar Physics (KIS) in Freiburg (Germany). ChroTel observes the H$\alpha$ line with a Lyot-type narrow-band filter at 656.2\,nm with a full-width-at-half-maximum (FWHM) of  $\Delta\lambda = 0.05$\,nm and a cadence of three minutes, whereby higher cadences of up to one minute are possible on special request. The recorded images have a size of $2048 \times 2048$ pixels. The flat-field and dark corrected H$\alpha$ filtergrams have been available online since April 2012\footnote{KIS Science Data Centre: \href{https://archive.sdc.leibniz-kis.de/}{archive.sdc.leibniz-kis.de}}. Technical problems resulted in larger data gaps of several months, that is in 2017 and 2019. For an overview of the acquired data between 2012 and 2018, readers can refer to Fig.~1 in \citet{Diercke2019b}. In Table~\ref{table:chrotel_data}, we present the number of observing days per year. The ChroTel observations cover the maximum of Solar Cycle~24 and the declining activity phase until the minimum of solar activity. Therefore, we have a representative statistical sample of the different activity phases within Solar Cycle~24.

\begin{table}[th]
\caption{Overview of H$\alpha$ observations with the Chromospheric Telescope in 2012\,--\,2020.}             % title of Table
\label{table:chrotel_data}      % is used to refer this table in the text
\centering                          % used for centering table
\begin{tabular}{c c}        % centered columns (2 columns)
\hline\hline                 % inserts double horizontal lines
Year & Number of observing days\rule[-6pt]{0pt}{18pt} \\    % table heading 
\hline                        % inserts single horizontal line
2012 & \rule{4pt}{0pt}146$^\dagger$\rule{0pt}{11pt}\\      % inserting body of the table
2013 & 210 \\
2014 & 172 \\
2015 & 115 \\
2016 & 102 \\
2017 & \rule{5pt}{0pt}76 \\
2018 & 141 \\
2019 & \rule{5pt}{0pt}38 \\
2020 & \rule{10pt}{0pt}56$^\star$\rule[-4pt]{0pt}{10pt} \\
\hline                                   %inserts single line
\end{tabular}
\parbox{50mm}{\smallskip
\footnotesize{$^\dagger$} starting 13 April \\
\footnotesize{$^\star$} until 2 September}
\end{table}

All images underwent basic data processing steps. We downloaded the flat-field and dark corrected H$\alpha$ level-1 data. To study long-term solar activity, we selected one H$\alpha$ filtergram per day by calculating the median filter-gradient similarity \citep[MFGS,][]{Deng2015, Denker2018}, where magnitude gradients of an input image and its median-filtered companion are compared. We selected the image with the highest MFGS value for each observing day, representing the image with the highest contrast. The selected filtergrams  were rotated so that solar north is on top and east is to the left, and the elliptical shape of the Sun was corrected for early morning and late evening data,  caused by differential refraction. All images were scaled so that the solar radius equals $r_\odot=1000$\,pixels, which results in an image scale of about 0\farcs96\,pixel$^{-1}$. Moreover, the Lyot filter introduced a nonuniform intensity pattern across the solar disk which was corrected by approximating the pattern with Zernike polynomials \citep{Denker2005, Shen2018}. All images were limb-darkening corrected, normalized to the median value of the quiet-Sun intensity $I_\mathrm{med}$, and aligned. Details of the data processing  are described in \citet{Diercke2019b}.

%-------------------------------------- Two column figure (place early!)
   \begin{SCfigure*}
   \includegraphics[width=0.67\textwidth]{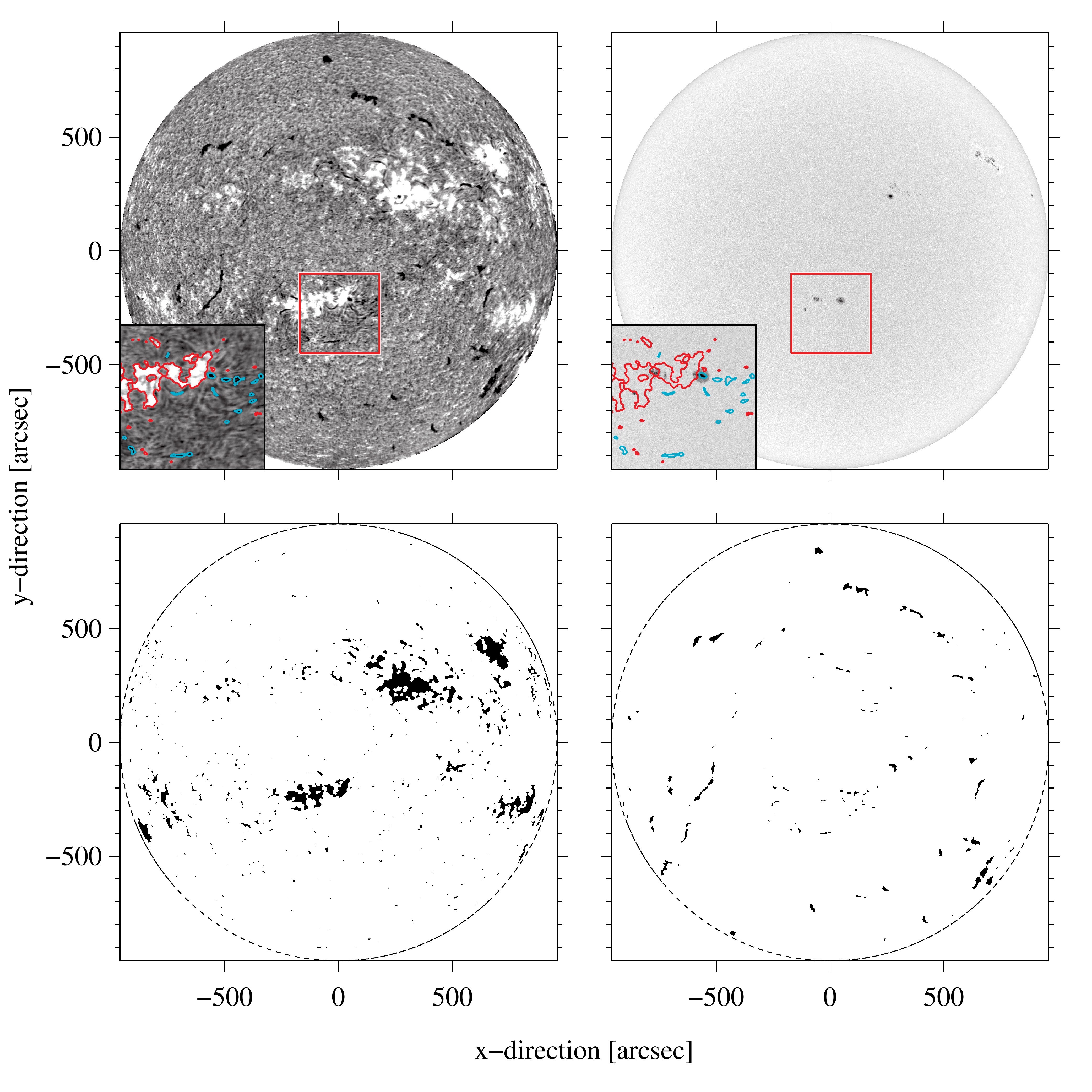}
   \caption{ChroTel H$\alpha$ filtergram on 2013~April~13 (top left) and the corresponding mask for bright (bottom left) and dark features (bottom right). For comparison, we display the corresponding continuum map from HMI/SDO (top right). The solar disk is indicated as a dashed circle. The red square indicates the location of the magnified region in the bottom left corner to visualize a sample region. The red and blue contours indicate the selected bright and dark features for the masks, respectively.}
    \label{Fig:method}%
    \end{SCfigure*}
%-----------------------------------------------------------------

\section{Method} \label{sec:method}

In order to explore the capabilities of H$\alpha$ as a proxy for solar activity, we use the H$\alpha$ excess calculated for bright regions and the H$\alpha$ deficit for dark (absorption) features. To extract bright regions, related to plage, we created a mask with the threshold $T^\mathrm{E}_{100}$ belonging to an intensity 10\% above the median intensity $I_\mathrm{med}$:
\begin{equation}
    T^\mathrm{E}_{100}= I_\mathrm{med} + 0.1\times I_\mathrm{med}.
\label{eq:thresh_exc}
\end{equation}
The threshold was selected so that the bright plage regions are isolated from the surrounding quiet-Sun.

The threshold $T^\mathrm{D}_{100}$ for the absorption features is defined with the following:
\begin{equation}
    T^\mathrm{D}_{100} = I_\mathrm{med} - 0.1\times I_\mathrm{med},
\label{eq:thresh_def}
\end{equation}
whereby we do not differentiate between sunspots and filaments. In Fig.~\ref{Fig:method} (top left), we display the original H$\alpha$ filtergram. The red box indicates the area magnified in the bottom left corner. The  red contours illustrate the isolated patches of plage and the blue contours show the selected dark absorption features, such as filaments and sunspots. Compared to continuum images from HMI/SDO (Fig.~\ref{Fig:method}, top right), some sunspots and pores are not visible in the H$\alpha$ filtergrams and, therefore, they are also not included in the masks. The mask for the bright plage regions for the entire solar disk is displayed in the bottom right panel and that for the dark absorption features is shown in the bottom right panel of Fig.~\ref{Fig:method}. Larger filaments are well recovered in the mask, but also small-scale filaments are recognized.

We used morphological image processing to extract these regions more efficiently. First, we applied morphological closing on the masks using a circular structuring element with a radius of ten pixels. We excluded small areas with less than 20\,pixels and removed border effects at 0.99\,$r_\odot$. From all pixels $R_{ij}$, we calculated the imaging H$\alpha$ excess $E_{100}$ of the bright regions similar to the method described in \citet{Johannesson1998} and \citet{Naqvi2010}. For the intensity $I_{ij}$ of each pixel, the following equation is defined:
\begin{equation}
    E_{100} = \frac{1}{1000} \sum_{ij} f_{ij} \quad \mathrm{with}\;
    \begin{cases}
    f_{ij} = I_{ij} - T^\mathrm{E}_{100} & \text{if } \ I_{ij} \geq T^\mathrm{E}_{100}\\
    f_{ij} = 0 & \text{if } \  I_{ij} < T^\mathrm{E}_{100}
  \end{cases}
\label{eq:excess}
.\end{equation}
The nomenclature is compliant to the definition in \citet{Naqvi2010}. Dividing by the factor 1000 results in values in a convenient data range.

%-------------------------------------- Two column figure (place early!)
   \begin{figure*}[th]
   \centering
   \includegraphics[width=\textwidth]{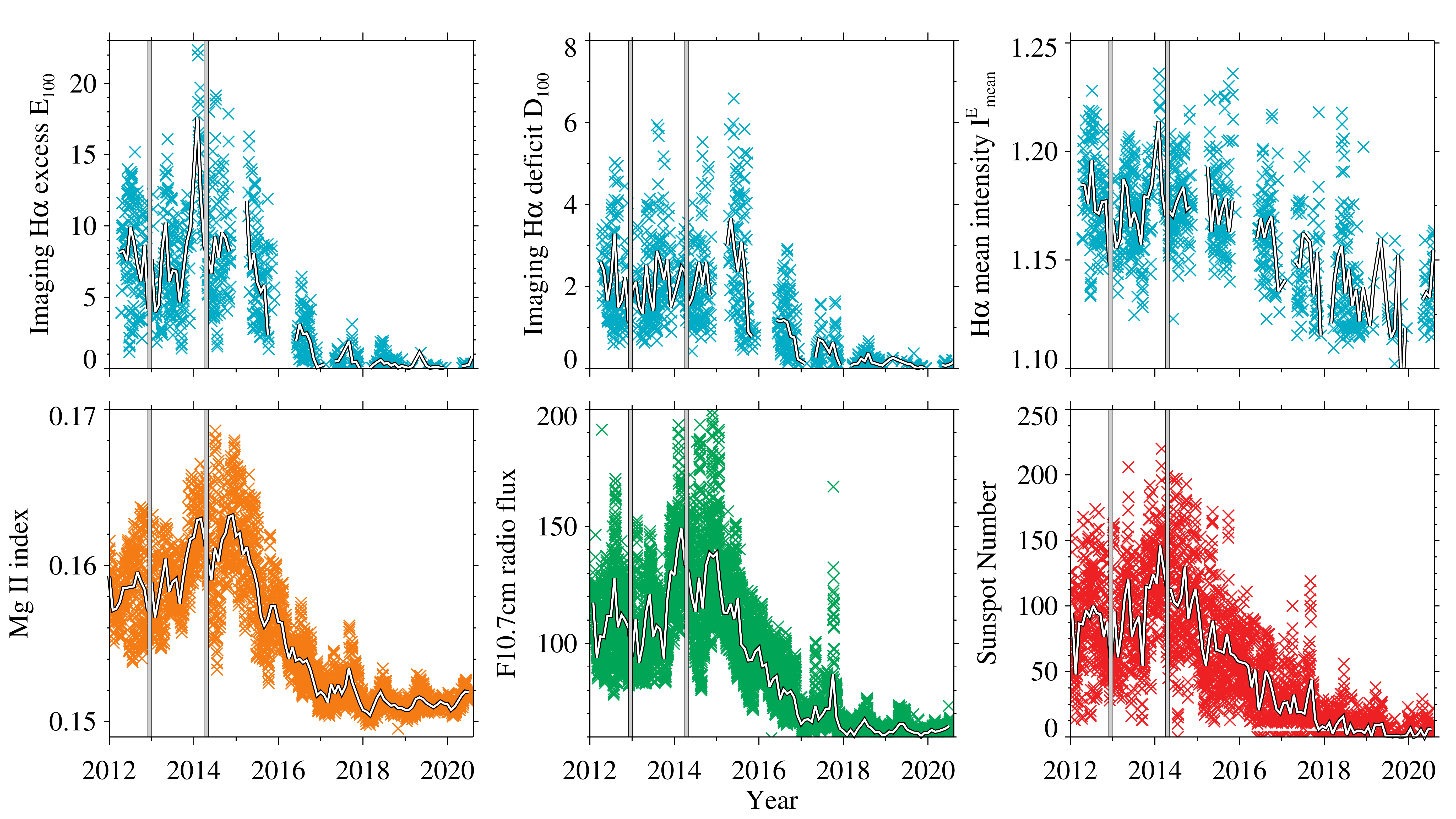}
   \caption{Comparison of the imaging H$\alpha$ excess with different solar activity indices during Solar Cycle~24 in 2012\,--\,2020. Top row: Imaging H$\alpha$ excess $E_{100}$ of bright features (left), imaging H$\alpha$ deficit $D_{100}$ of dark features (middle), and mean intensity of H$\alpha$ excess regions $I^\mathrm{E}_\mathrm{mean}$ (right). Bottom row: \mbox{Mg\,\textsc{ii}} index (left), F10.7\,cm radio flux (middle), and relative sunspot number (right) for daily observations. The monthly average for all tracers is shown as a white line. The gray areas in the background indicate the northern and southern magnetic field reversals in November 2012 and March 2014, respectively.}
    \label{Fig:index_comp}%
    \end{figure*}
%-----------------------------------------------------------------

As we defined the imaging H$\alpha$ excess, we define the imaging H$\alpha$  deficit $ D_{100}$ from the threshold $T^\mathrm{D}_{100}$ for the following dark absorption features:
\begin{equation}
    D_{100} = \frac{1}{1000} \sum_{ij} f_{ij} \quad \mathrm{with}\;
    \begin{cases}
    f_{ij} = |I_{ij} - T^\mathrm{D}_{100}| & \text{if } \ I_{ij} \leq T^\mathrm{D}_{100}\\
    f_{ij} = 0 & \text{if } \  I_{ij} > T^\mathrm{D}_{100}
  \end{cases}
\label{eq:deficit}
.\end{equation}
In Section~\ref{sec:halpha_index}, we compare the imaging H$\alpha$ excess and deficit with the mean intensity $I^\mathrm{E}_\mathrm{mean}$ of all bright regions with an intensity $I_{ij}$ above threshold $T^\mathrm{E}_{100}$:
\begin{equation}
    I^\mathrm{E}_\mathrm{mean} = \frac{1}{n_\mathrm{T}} \sum_{ij} f_{ij} \quad \mathrm{with}\;
    \begin{cases}
    f_{ij} = I_{ij} & I_{ij} \geq T^\mathrm{E}_{100}\\
    f_{ij} = 0 & I_{ij} < T^\mathrm{E}_{100}
  \end{cases}
 \label{eq:mean}
.\end{equation}
The mean intensity $I^\mathrm{E}_\mathrm{mean}$ is the sum of all intensities $f_\mathrm{ij}$  above threshold $T^\mathrm{E}_{100}$ divided by the number of pixels $n_\mathrm{T}$ above the threshold. In relation to this, we define the mean intensity of dark absorption features $I^\mathrm{D}_\mathrm{mean}$ equivalent to $I^\mathrm{E}_\mathrm{mean}$ with $I_{ij} \leq T^\mathrm{D}_{100}$. Both the mean excess intensity $I^\mathrm{E}_\mathrm{mean}$ and the mean deficit intensity $I^\mathrm{D}_\mathrm{mean}$ are compared relative to the median intensity $I_\mathrm{med}$.

Another quantity we want to introduce here is the mean excess intensity $I^\mathrm{E}_\mathrm{mean, K}$ of individual isolated patches K in an image. The definition is equivalent to Equation~(\ref{eq:mean}), but calculated for each individual patch separately. This quantity is used in Sect.~\ref{sec:int_dist}, where we calculate their distribution for a certain time span during minimum and maximum.

Furthermore, we define the area coverage fraction of the H$\alpha$ excess $A_\mathrm{E}$, which is the total number of pixels $n_\mathrm{T}$ above threshold $T^\mathrm{E}_{100}$ divided by the total number of pixels belonging to the solar surface in the filtergram $P_\mathrm{N}$:
\begin{equation}
    A_\mathrm{E} = \frac{n_\mathrm{T}}{100 \times P_\mathrm{N}} 
.\end{equation}
The area coverage fraction $A_\mathrm{E}$ is used in Section~\ref{sec:index_rotation}.The area coverage fraction of the H$\alpha$ deficit $A_\mathrm{D}$ is defined in the same way. In Section~\ref{sec:index_ar}, we use only the total number of pixels $n_\mathrm{T, AR}$ above threshold $T^\mathrm{E}_{100}$, which belong to selected active regions (ARs). Here, we inspect the H$\alpha$ excess regions of the active region, which are a collection of several isolated patches of plage.

Several established solar activity tracers are commonly used in the solar community. To study the value of the H$\alpha$ excess and deficit as a tracer of solar activity, in the following  we compare both quantities with three established proxies: spectral magnesium \mbox{Mg\,\textsc{ii}} index,\footnote{\texttt{\href{https://www.iup.uni-bremen.de/gome/gomemgii.html}{https://www.iup.uni-bremen.de}}} the solar radio flux at F10.7\,cm,\footnote{\texttt{\href{https://spaceweather.gc.ca/solarflux/sx-5-en.php}{https://spaceweather.gc.ca}}} and the sunspot number.\footnote{\texttt{\href{http://www.sidc.be/silso/datafiles}{http://www.sidc.be}}} For each index, we use the daily value and the monthly average value. For the imaging H$\alpha$ excess and deficit, we calculate the monthly average for the complete time period.  To compare the H$\alpha$ excess and deficit with the other tracers, we calculate their  monthly average. The Spearman  correlation $\rho_\mathrm{S}$ is then calculated for the monthly average of each tracer.

\section{Results} \label{sec:results}

\subsection{Imaging H\texorpdfstring{$\alpha$}{-alpha} excess and deficit} \label{sec:halpha_index}

In this section, we analyze the imaging H$\alpha$ excess and deficit for bright and dark regions during Solar Cycle~24. We compare both quantities to the mean intensity of the bright regions $I^\mathrm{E}_\mathrm{mean}$, as well as to other activity indices, that is to say the relative sunspot number, the F10.7\,cm radio flux, and the \mbox{Mg\,\textsc{ii}} index. The motivation of this study is to characterize the intensity excess from the bright regions for possible relations to stellar activity studies. The H$\alpha$ deficit might help the solar-stellar connection community to quantify how much the H$\alpha$ intensity decreases across the solar or stellar disk because of absorption features in the atmosphere.

Solar Cycle~24 started in late 2008, but ChroTel observations have been available in the archive since April 2012. The maximum of Solar Cycle~24 on the northern hemisphere was in November 2012, which was followed by the maximum on the southern hemisphere about 16 months later in March 2014 \citep{Sun2015}. The recorded data cover the activity minimum in 2018 and 2019, as well as the start of Solar Cycle~25 in 2019 and 2020. 

In Figure~\ref{Fig:index_comp}, we compare the H$\alpha$ excess $E_{100}$ (top left panel) directly with the  H$\alpha$ deficit  $D_{100}$ of the absorption features  (top middle panel). Both indices are set against the mean intensity $I^\mathrm{E}_\mathrm{mean}$ of H$\alpha$ excess regions, as described in Equation~(\ref{eq:mean}),  throughout Solar Cycle~24 (top right panel).  The daily values (cyan x) as well as the monthly average (white curve) are displayed. Comparing the H$\alpha$ deficit to the H$\alpha$ excess, both curves show a cyclic behavior related to the solar activity.  The H$\alpha$ excess is highest around maximum and decreases toward the minimum when the number of active regions and the related bright plage regions is lower (Fig.~\ref{Fig:index_comp}, top left panel). The curve of the H$\alpha$ deficit for the absorption features (top middle panel in Fig.~\ref{Fig:index_comp}) follows the global evolution of the solar activity, as well, but it peaks at different times. The vertical gray bars in Fig.~\ref{Fig:index_comp} indicate the magnetic field reversal in the northern (left panel) and southern hemisphere (right panel) during the maximum of Solar Cycle~24 \citep{Sun2015}. The H$\alpha$ excess has its highest peak around the time of the magnetic field reversal in the southern hemisphere (second gray bar) in 2014, but with smaller peaks before and after the first magnetic field reversal in the northern hemisphere (first gray bar). The H$\alpha$ deficit has one peak before each magnetic field reversal related to each hemisphere, but at a later time as for the H$\alpha$ excess. The highest peak is around the end of 2015. Afterwards the H$\alpha$ deficit reduces to a minimum, similar to the H$\alpha$ excess. Nonetheless, the correlation of the monthly averaged  H$\alpha$ excess and deficit is high with a Spearman correlation of $\rho_\mathrm{ED} = 0.91$.

The changes between the maximum and minimum of the solar cycle are best visible in the H$\alpha$ excess $E_{100}$. Here the values vary between $E^{\mathrm{max}}_{100} \approx 20$ around the activity maximum and $E_{100} \approx 1$ around the activity minimum. The spread of the daily H$\alpha$ excess around the maximum varies between $E_{100} \approx 5$ and $E_{100} \approx 20$, whereas around the minimum the variations are much smaller between $E_{100} \approx 2$ and $E_{100} \approx 0$.  This is related to the very small number of active regions during the minimum. We can see that the  H$\alpha$ excess and deficit increase shortly before the magnetic field reversal in the northern hemisphere (left gray bar). For the magnetic field reversal in the southern hemisphere (right gray bar), the H$\alpha$ excess increases just before the magnetic field reversal to its maximum and it stays high for several months.

For the H$\alpha$ deficit of absorption features, the values vary between $D_{100} \approx 7$ and $D_{100} \approx 0$  between maximum and minimum, respectively. The variation is much stronger in the maximum, covering the full range from maximum values to low values of about $D_{100} \approx 1$ in just a few days, while variations of the daily H$\alpha$ deficit are very small during the minimum. Just before the magnetic field reversal in the northern and southern hemispheres, the  H$\alpha$ deficit is high and drops after the reversal. Nonetheless, the number of absorption structures increases to an even higher level after the magnetic field reversal. The drop from a high number of absorption features toward the minimum is steeper than for the bright regions. We discuss the H$\alpha$ deficit in detail in Section~\ref{sec:disc}.

The variations of the mean intensity $I^\mathrm{E}_\mathrm{mean}$ of H$\alpha$ excess regions  (top right panel in Fig.~\ref{Fig:index_comp}) are large throughout the entire cycle, but decrease toward the minimum. The daily variations decrease in the minimum and the intensity level is clearly reduced. Comparing the mean intensity of H$\alpha$ excess regions to the imaging H$\alpha$ excess, we recognize that the H$\alpha$ excess is better suited to reflect the solar activity than the mean intensity, because in the mean intensity, the daily variations are much larger. Especially the declining phase of the solar cycle is better represented by the imaging H$\alpha$ excess. The H$\alpha$ excess depends on the number of pixels, which go into the calculation, resulting in a better scaling of the curve and a better comparison to other activity tracers.

In the following, we compare the H$\alpha$ excess and deficit with the spectral magnesium \mbox{Mg\,\textsc{ii}} index, the solar radio flux at F10.7\,cm,  and the sunspot number (lower panels in Fig.~\ref{Fig:index_comp}). We display the daily observations for each activity tracer as single points and their monthly average as a solid white line. The trend of the H$\alpha$ excess is very similar to the other three proxies throughout the solar cycle. Toward the maximum the daily variations of the H$\alpha$ excess are high, but in the monthly average the H$\alpha$ excess is much higher than in the minimum, where the daily variation is also reduced. The first peak of the H$\alpha$ excess is due to the maximum in the northern hemisphere. The same peak is visible in the solar radio flux at F10.7\,cm, the sunspot number, and it is slightly shifted in the \mbox{Mg\,\textsc{ii}} index as well. The second peak in the H$\alpha$ excess coincides with a peak in the sunspot number and a small peak in the F10.7\,cm radio flux. The high increase in the excess in H$\alpha$ shortly before the magnetic field reversal in the southern hemisphere (right gray bar) is similar to the sunspot number and the F10.7\,cm radio flux as well, whereby the latter stays at a high level for several month after the reversal, whereas the H$\alpha$ excess decreases in a similar manner as the \mbox{Mg\,\textsc{ii}} index. The maximum of the \mbox{Mg\,\textsc{ii}} index is shortly after the magnetic field reversal in the southern hemisphere. There is another increase in the H$\alpha$ excess in the declining phase of the cycle in the end of 2017, which is also visible in all the other three tracers at the same time. The correlation of the  H$\alpha$ excess is lowest for the smoothed sunspot number with $\rho_\mathrm{E,SSN} = 0.88$ and $\rho_\mathrm{E, mSSN} = 0.91$ with the mean monthly sunspot number. The correlation to the F10.7\,cm radio flux is $\rho_\mathrm{E, F10} = 0.93$. The highest correlation is with the \mbox{Mg\,\textsc{ii}} index  with a correlation of  $\rho_\mathrm{E, Mg} = 0.95$. The H$\alpha$ excess and the \mbox{Mg\,\textsc{ii}} index display both the solar activity of the chromosphere. 

The first peak of the H$\alpha$ deficit coincides with the three other tracers. Their is a smaller peak after the magnetic field reversal of the southern hemisphere, which coincides with the maximum of the  \mbox{Mg\,\textsc{ii}} index and the F10.7\,cm radio flux, but this is most probably just an artificial peak due to a lack of data in the beginning of 2015. The real peak of the H$\alpha$ deficit is at the end of 2015, when most other tracers have already started to decline. The additional peak toward the minimum in the end of 2017 is visible for the H$\alpha$ deficit as well. The Spearman correlation of the monthly mean of the H$\alpha$ deficit with the other three tracers of solar activity is lower compared to the H$\alpha$ excess, but still relatively high. The highest correlation was calculated between the H$\alpha$ deficit and the F10.7cm radio flux with  $\rho_\mathrm{D, F10} = 0.89$, followed by the correlation with the \mbox{Mg\,\textsc{ii}} index  $\rho_\mathrm{D, Mg} = 0.87$. The correlation with the smoothed sunspot number and the  mean monthly sunspot number is $\rho_\mathrm{D,SSN} = 0.85$ and $\rho_\mathrm{D, mSSN} = 0.84$.

%-------------------------------------- two column figure
   \begin{figure}[t]
   \centering
   \includegraphics[width=\columnwidth]{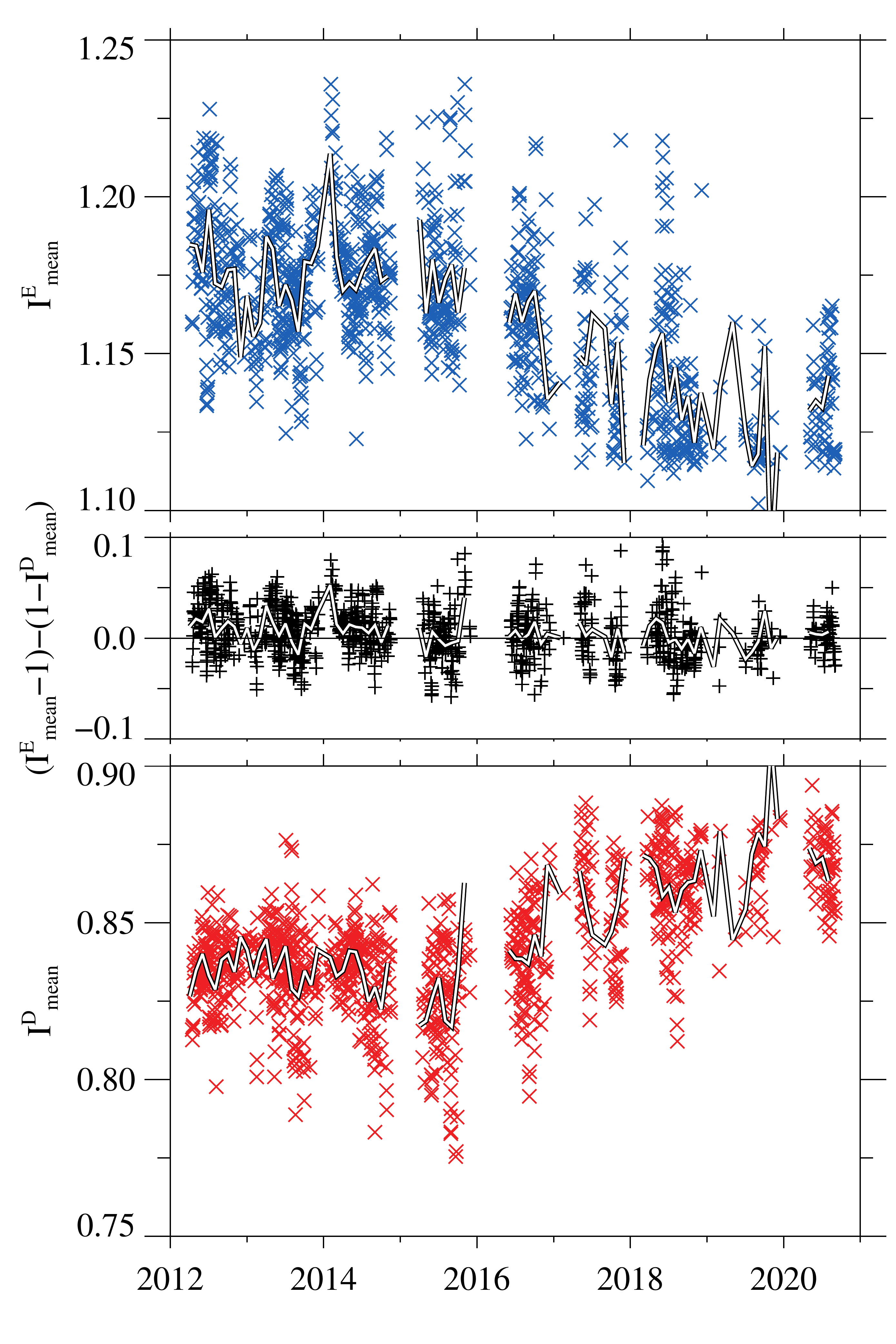}
   \caption{Comparison of the mean intensity of bright H$\alpha$ regions $I^\mathrm{E}_\mathrm{mean}$ (blue crosses, top panel) and the mean intensity of dark H$\alpha$ absorbing regions $I^\mathrm{D}_\mathrm{mean}$ (red crosses, bottom panel) relative to the median intensity. In the middle panel, the relative distance to an intensity of $I_\mathrm{mean} = 1$ of $I^\mathrm{E}_\mathrm{mean}$ and $I^\mathrm{D}_\mathrm{mean}$ is displayed. In addition, the monthly mean is shown with a white line.}
    \label{Fig:mean_compare}%
    \end{figure}
%-----------------------------------------------------------------

In the following section, we compare the mean intensity of bright H$\alpha$ regions $I^\mathrm{E}_\mathrm{mean}$ and the mean intensity of dark H$\alpha$ absorbing regions $I^\mathrm{D}_\mathrm{mean}$. Additionally, for both quantities, we calculate their difference to the median intensity $I_\mathrm{med}$. All images are normalized to the median intensity, which means that $I_\mathrm{med} \approx 1$. Together with their monthly mean, these quantities are displayed in Fig.~\ref{Fig:mean_compare}. The behavior of $I^\mathrm{E}_\mathrm{mean}$ and $I^\mathrm{D}_\mathrm{mean}$ throughout the years is very similar. During the maximum, $I^\mathrm{E}_\mathrm{mean}$ is high indicating a large number of bright regions and $I^\mathrm{D}_\mathrm{mean}$ is small, indicating a larger number of absorbing features on the Sun. Toward the minimum, $I^\mathrm{E}_\mathrm{mean}$ and $I^\mathrm{D}_\mathrm{mean}$  are decreasing and increasing, respectively, approaching values closer to  $I_\mathrm{med}$. Sighting the relative difference of both mean intensities, the difference is close to zero throughout the observing period. The Spearman correlation of the monthly averaged mean intensity for the H$\alpha$ excess and deficit is high with a Pearson anticorrelation of $\rho_\mathrm{E, D} = -0.81$.

\subsection{Intensity distribution of active regions} \label{sec:int_dist}

\begin{figure}[t]
   \centering
   \includegraphics[width=\columnwidth]{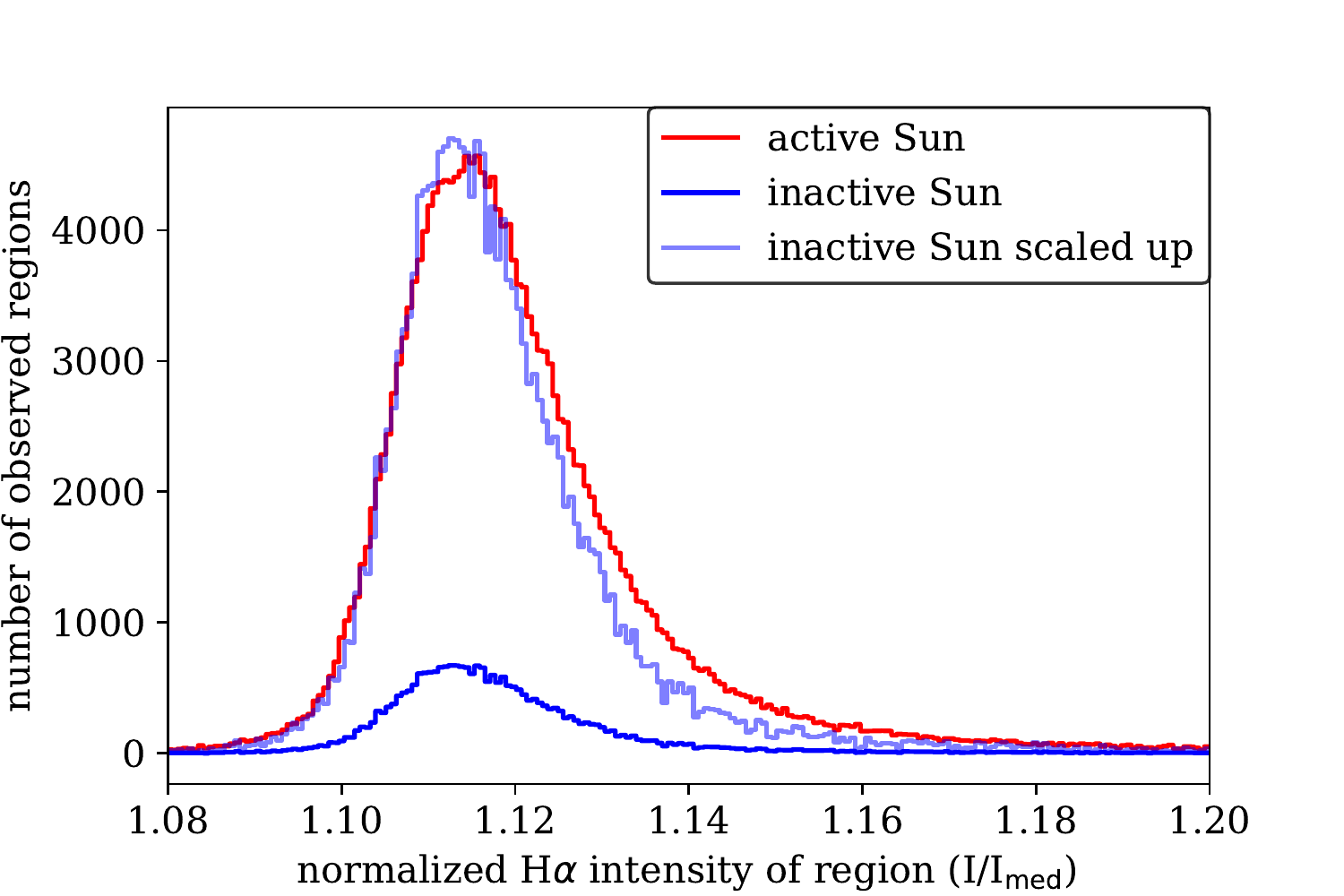}
   \caption{Intensity distribution of $I^\mathrm{E}_\mathrm{mean, K}$ for individual regions above the threshold  shown as a histogram for the active Sun (red) and the inactive Sun (blue). A scaled-up histogram of the inactive Sun is provided for an easier visual comparison with the active Sun (light blue).}
    \label{Fig:intensity_distribution}%
    \end{figure}

In the context of solar and stellar activity, not only is the number of active regions of interest, but as is the distribution of H$\alpha$ intensities of the observed regions. To investigate how the active (during solar maximum) and inactive parts (during solar minimum) of the solar cycle differ in the H$\alpha$ properties of active regions, we selected individual isolated patches of H$\alpha$ excess regions observed on the Sun solar minimum and maximum, as described in Sect.~\ref{sec:method}. For the solar maximum, we used observations between 2012 and 2015, which include 202\,991 isolated patches of H$\alpha$ excess regions in total. For the solar minimum, we used data between 2016 and 2020 including 25\,426 isolated patches. The number of active regions during the solar activity minimum is very low, which was the reason why we decided to calculate the mean intensity $I^\mathrm{E}_\mathrm{mean, K}$ of individual isolated patches of H$\alpha$ excess. We produced a histogram of the number of observed isolated patches as a function of the normalized H$\alpha$ mean intensity (Fig.~\ref{Fig:intensity_distribution}). The total number of bright regions between solar maximum (red histogram) and minimum (blue histogram) differs by a factor of about four. The shape of those two distributions is almost the same, as can be seen when scaling the distribution for the inactive Sun up to the level of the active Sun (light blue histogram). Both distributions resemble a skewed Gaussian, with a longer tail toward higher intensities. The tail is slightly more extended for the active Sun. The application of this result is discussed in Sect.~\ref{sec:disc_E100}.

%-------------------------------------- two column figure
%-----------------------------------------------------------------

%-------------------------------------- One column figure
   \begin{figure*}[th]
   \centering
   \includegraphics[width=\textwidth]{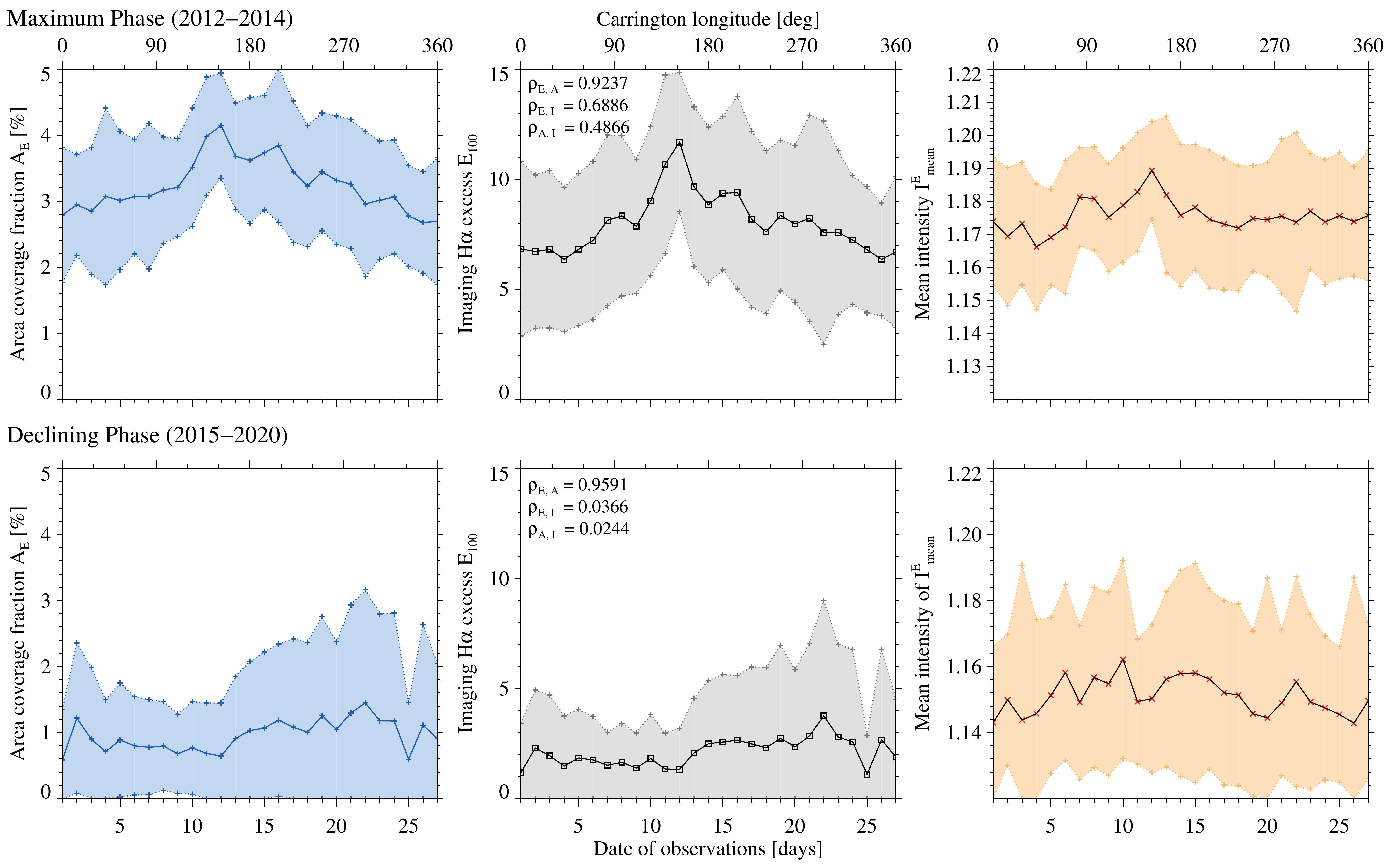}
   \caption{Area coverage fraction of H$\alpha$ excess regions in percent (left row), the mean intensity of the H$\alpha$ excess regions (right row), and the imaging H$\alpha$ excess $E_{100}$ (middle row) averaged for all Carrington rotations during the maximum phase of Solar Cycle~24 (upper column) and the declining phase of Solar Cycle~24 (lower column). The shaded areas indicate the standard deviation. In the middle row, we provide the Spearman correlation coefficient  $\rho_\mathrm{E, A}$ of the H$\alpha$ excess and the area coverage fraction, the correlation coefficient $\rho_\mathrm{E, I}$ of the H$\alpha$ excess and the mean intensity, and the correlation coefficient of the area coverage fraction and the mean intensity $\rho_\mathrm{A, I}$.} 
    \label{Fig:coverage_all}%
    \end{figure*}
%-----------------------------------------------------------------

%-------------------------------------- One column figure
   \begin{figure*}[th]
   \centering
   \includegraphics[width=\textwidth]{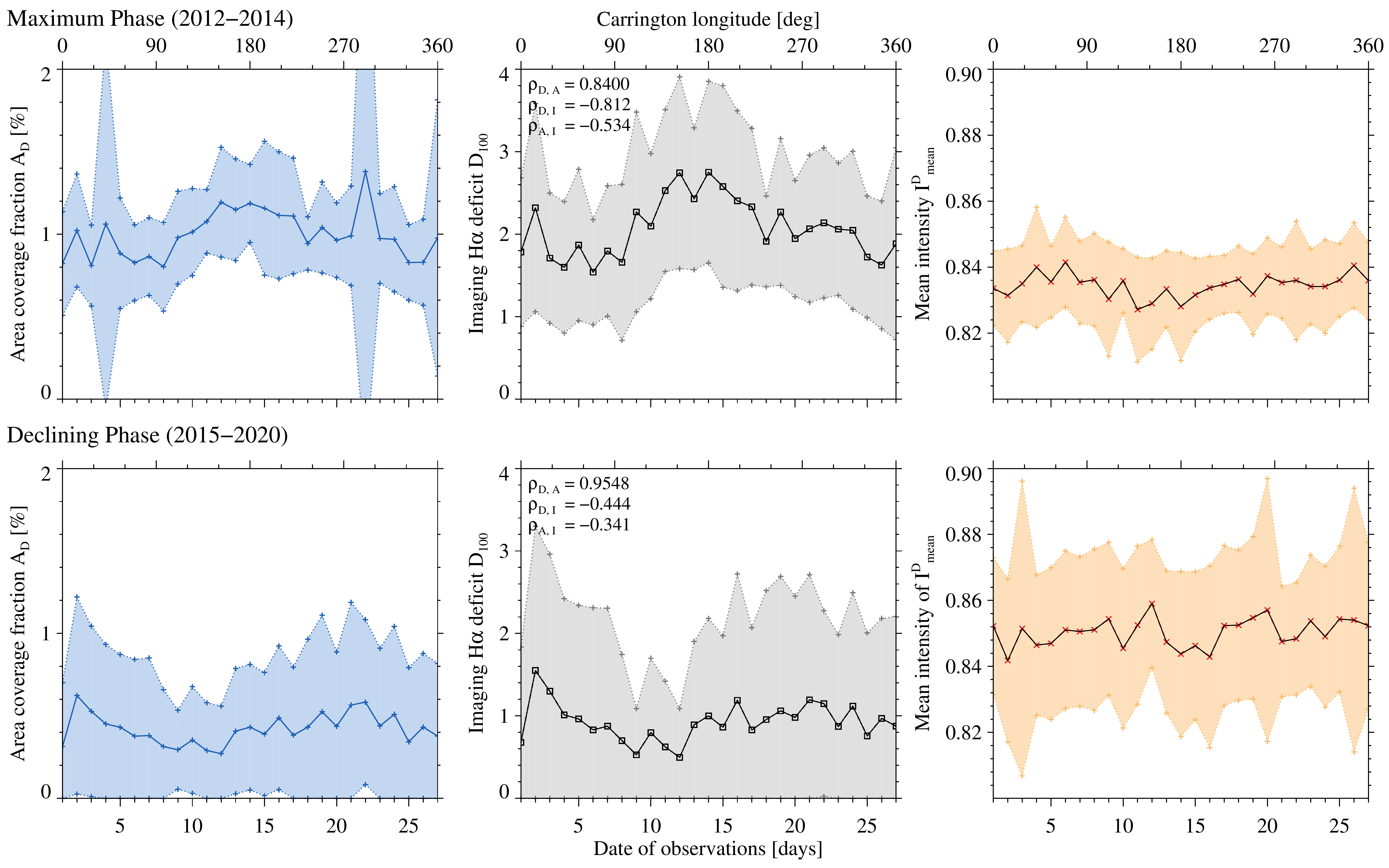}
   \caption{Area coverage fraction of H$\alpha$ deficit regions in percent (left row), the mean intensity of the H$\alpha$ deficit regions (right row), and the imaging H$\alpha$ deficit $D_{100}$ (middle row) averaged for all Carrington rotations during the maximum phase of Solar Cycle~24 (upper column) and the declining phase of Solar Cycle~24 (lower column). The shaded areas indicate the standard deviation. In the middle row, we provide the Spearman correlation coefficient  $\rho_\mathrm{D, A}$ of the H$\alpha$ deficit and the area coverage fraction, the correlation coefficient $\rho_\mathrm{D, I}$ of the H$\alpha$ deficit and the mean intensity, and the correlation coefficient of the area coverage fraction and the mean intensity $\rho_\mathrm{A, I}$.} 
    \label{Fig:coverage_rev_all}%
    \end{figure*}
%-----------------------------------------------------------------

\subsection{Active longitude of the \texorpdfstring{H$\alpha$}{-alpha} excess and deficit}
\label{sec:index_rotation}

In this section, we inspect solar rotation for all Carrington rotations in the data set in detail, separating the data set into maximum (2012\,--\,2014) and declining phase (2015\,--\,2020) for both the H$\alpha$ excess and deficit (Figs.~\ref{Fig:coverage_all} and ~\ref{Fig:coverage_rev_all}, respectively). For the Sun, a Carrington rotation is defined with a rotation period of 27.2753\,days \citep{Knaack2004}. The ChroTel data set contains 84 Carrington rotations. We sorted each observed image in its respective Carrington frame and calculated the average of the area coverage fraction $A_\mathrm{E}$, the imaging H$\alpha$ excess $E_{100}$ and deficit $D_{100}$, and the mean intensity $I^\mathrm{E}_\mathrm{mean}$ (Fig.~\ref{Fig:coverage_all}). For each day of an average Carrington rotation, between 13 and 24 data points are summed up. We calculated the standard deviation  for each quantity. The same was repeated for the H$\alpha$ deficit and their respective quantities (Fig.~\ref{Fig:coverage_rev_all}).

In Fig.~\ref{Fig:coverage_all}, the mean area coverage fraction of the H$\alpha$ excess, the mean H$\alpha$ excess, and the mean intensity averaged for one Carrington period are displayed. During the maximum phase (upper left and middle panel), there is a clear bell shape of the mean area coverage fraction and the H$\alpha$ excess with the maximum in the middle of the Carrington reference frame, indicating the active longitude of Solar Cycle~24 at about 150\arcdeg. For the mean intensity, the bell shape is not so clearly visible, but still at around the middle of the Carrington rotation, the mean intensity is slightly higher. The Spearman correlation between the area coverage fraction and the H$\alpha$ excess is very high with $\rho_\mathrm{E,A} = 0.92$, but also the Spearman correlation between the H$\alpha$ excess and mean intensity is relatively high with $\rho_\mathrm{E,I} = 0.69$. On the other hand, the Spearman correlation of the area coverage fraction and the mean intensity is only $\rho_\mathrm{A,I} = 0.49$.

In the declining phase (Fig.~\ref{Fig:coverage_all}, lower panel), the bell shape completely disappeared for all three quantities. The area coverage fraction reduced from values between 2\% and 5\% to values between 0\% and 3\%. The imaging  H$\alpha$ excess reduced from values between 4 and 15 to values between 0 and 10. The mean intensity still occasionally reaches values of up to 1.21\,$I/I_\mathrm{med}$, but lower intensities of about 1.13\,$I/I_\mathrm{med}$ are now also possible. The standard deviation for the mean intensity reaches higher values than during the maximum phase. The correlation is even higher for the area coverage fraction and the H$\alpha$ excess with  $\rho_\mathrm{E,A} = 0.96$, whereby the correlation with the intensity is very low ($\rho_\mathrm{E,I} = 0.03$). The same applies for the correlation of the area coverage fraction and the mean intensity  ($\rho_\mathrm{A,I} = 0.02$).

In Fig.~\ref{Fig:coverage_rev_all}, the averaged Carrington rotation for the imaging H$\alpha$ deficit, the covarage fraction $A_\mathrm{D}$ of the H$\alpha$ deficit, and the mean intensity of absorption features $I^\mathrm{D}_\mathrm{mean}$ is displayed. In the maximum phase (upper panel), the bell shape for the area coverage fraction and H$\alpha$ deficit is again visible, but the mean intensity does not visibly change during the averaged Carrington period. The peak of the H$\alpha$ deficit is between 150\arcdeg and 180\arcdeg in the Carrington reference frame, which is comparable to the peak of the H$\alpha$ excess. For the area coverage fraction, there are two very large spikes for the standard deviation, which may be correlated with very large filaments on the solar surface, which influence the standard deviation. In the declining  phase (lower panel), the bell shape is no longer visible for the coverage fraction and H$\alpha$ deficit. The values decrease for the area coverage fraction from about 0.8\% to 1.5\% to values below 0.8\%. The imaging H$\alpha$ deficit is on average between 1 and 3 in the maximum phase and it reduces to values between 0 and 2, which indicates that fewer filaments are present on the solar disk. The mean intensity slightly increases from about 0.84 to 0.85 on average, indicating fewer absorption features. The standard deviation visibly increases as well for the mean intensity during the declining phase, whereas the standard deviation is relatively low in the maximum phase. The correlation of the area coverage fraction and the H$\alpha$ deficit is lower in the maximum phase compared to the H$\alpha$ excess with $\rho_\mathrm{D,A} = 0.84$, but it increases to $\rho_\mathrm{D,A} = 0.95$ in the declining phase. The Spearman correlation for the H$\alpha$ deficit and the intensity indicates an anticorrelation $\rho_\mathrm{D,I} = -0.81$, but comparing both quantities, they do not seem to correlate at all. For the declining phase, only a correlation of $\rho_\mathrm{D,I} = -0.44$ was calculated. The correlation of the area coverage fraction and the mean intensity is in both the maximum phase and the declining phase which are relatively low with values of $\rho_\mathrm{A,I} = -0.53$ and $\rho_\mathrm{A,I} = -0.34$, respectively. The shape of the H$\alpha$ excess and deficit for an averaged Carrington rotation is discussed in Section~\ref{sec:disc_E100}.

\subsection{H\texorpdfstring{$\alpha$}{-alpha} excess of active regions} \label{sec:index_ar}

In this section, we discuss the variation of the mean intensity and H$\alpha$ excess during the passage of 15 different active regions between 2012 and 2020. The selected regions are listed in Table~\ref{table:chrotel_ar}. In the table, we marked the active regions with low activity (small number of C-class flares) and high activity (large number of C-class flares and at least one M-class flare) with asterisks.

The active region in July 2018 did not develop sunspots and therefore it did not receive a NOAA identification number, but because of the large, clear visibility in EUV images, this active region received an identification number with the Spatial Possibilities Clustering Algorithm \citep[SPoCA,][]{Verbeeck2014}. Since the region in the H$\alpha$ filtergrams was dominated by plage regions, we decided to include this region to our data sample.

%-------------------------------------- Two column figure (place early!)
   \begin{figure*}[th]
   \centering
   \includegraphics[width=\textwidth]{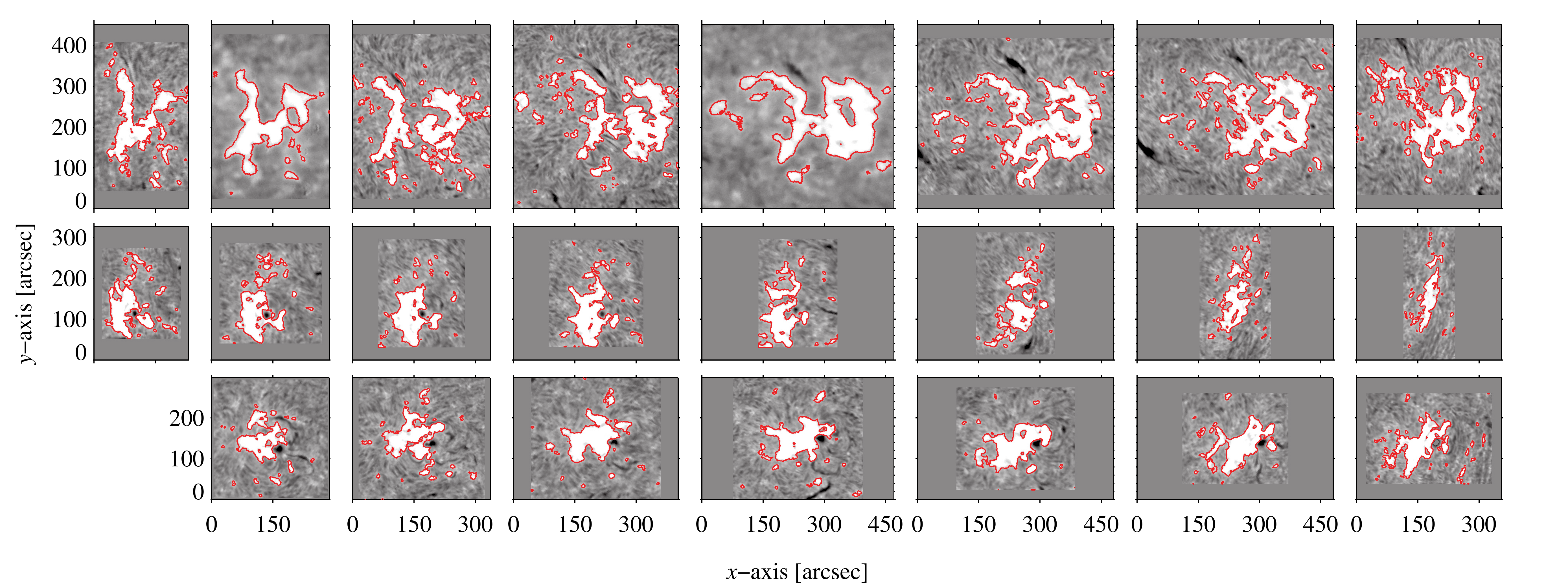}
   \caption{Regions of interest (ROIs) extracted from ChroTel H$\alpha$ full-disk filtergrams, displaying the disk passage (days 4\,--\,11) of three active regions in 2013 during the maximum of Solar Cycle~24: NOAA~11850 (top), NOAA~11835 (middle), and NOAA~11818 (bottom). The respective mask for each ROI is displayed with red contours.}
    \label{Fig:ar_overview}%
    \end{figure*}
%-----------------------------------------------------------------

\begin{table}[th]
\caption{Overview of selected active regions and their observing period between 2012\,--\,2020. The asterisk indicates the flaring activity of the active region.}           % title of Table
\label{table:chrotel_ar}      % is used to refer this table in the text
\centering                          % used for centering table
\begin{tabular}{cc}        % centered columns (2 columns)
\hline\hline                 % inserts double horizontal lines
NOAA & Observing period \rule[-6pt]{0pt}{18pt} \\    % table heading 
\hline                        % inserts single horizontal line
\rule{10pt}{0pt}11476$^{\star\star}$ & 2012/05/09 -- 2012/05/16\rule{0pt}{11pt}\\      % inserting body of the table
11555                                & 2012/08/26 -- 2012/09/04 \\
\rule{10pt}{0pt}11818$^{\star\star}$ & 2013/08/10 -- 2013/08/20  \\
11835                                & 2013/08/25 -- 2013/09/05  \\
\rule{5pt}{0pt}11850$^\star$         & 2013/09/19 -- 2013/10/01  \\
\rule{5pt}{0pt}12049$^\star$         & 2014/04/29 -- 2014/05/08  \\
\rule{5pt}{0pt}12139$^\star$         & 2014/08/11 -- 2014/08/22  \\
\rule{5pt}{0pt}12389$^\star$         & 2015/07/24 -- 2015/08/03  \\
12578                                & 2016/08/15 -- 2016/08/26 \\
12686                                & 2017/10/24 -- 2017/11/01 \\
12709                                & 2018/05/08 -- 2018/05/19  \\
\rule{5pt}{0pt}12712$^\star$         & 2018/05/24 -- 2018/06/04  \\
SPoCA 21904$^\dagger$                & 2018/07/08 -- 2018/07/18 \\
12767                                & 2020/07/22 -- 2020/08/01  \\
12769                                & 2020/08/01 -- 2020/08/13\rule[-4pt]{0pt}{10pt} \\
\hline                                   %inserts single line
\end{tabular}
\parbox{50mm}{\smallskip
\footnotesize{$^\dagger$}\rule{5pt}{0pt} NOAA identification not available\\
\footnotesize{$^\star$}\rule{5pt}{0pt} Flaring active regions (C-class)\\
\footnotesize{$^{\star\star}$} Flaring active regions (M-class)
}
\vspace{-0.5cm}
\end{table}

\begin{figure*}[th]
   \centering
   \includegraphics[width=\textwidth]{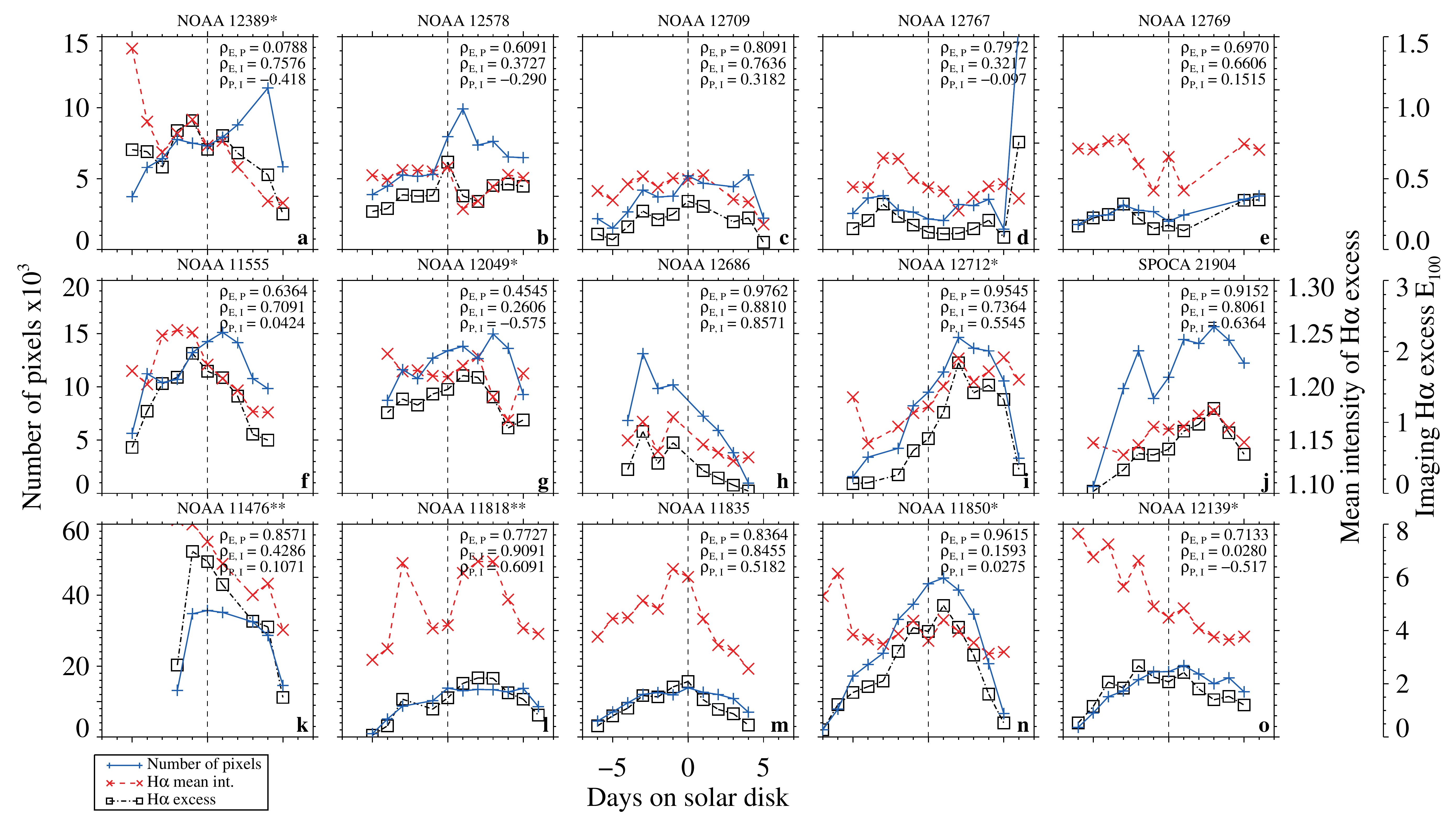}
   \caption{H$\alpha$ excess (dash-dotted black line), mean intensity (dashed red line), and number of pixels in mask (solid blue line) for 15 active regions during their disk passage. The vertical dashed lines indicate the crossing of the the active region through the central meridian. For each active region, we calculated the Spearman correlation of the H$\alpha$ excess with the number of pixels $\rho_\mathrm{E, P}$ and with the intensity $\rho_\mathrm{E, I}$, as well as the Spearman correlation of the number of pixels and the intensity $\rho_\mathrm{P, I}$. The asterisks next to the active region number indicate the flaring activity (see Table~\ref{table:chrotel_ar}).}
    \label{Fig:mean_int_ar_all}%
    \end{figure*}

We selected active regions NOAA~11818 (2013 August 9\,--\,20, 12~days), NOAA~11835 (2013 August~25 -- September~4, 11 days), and NOAA~11850 (2013 September~19 -- October~1, 13 days) to illustrate their disk passage in the H$\alpha$ filtergrams (Fig.~\ref{Fig:ar_overview}). For the sake of clarity, we concentrated on the eight central days (days 4\,--\,11) and discarded the limb observations. The active regions NOAA~11850 and NOAA~11818 showed flux emergence until about the middle of their disk passage and they decayed in the following days. The regions developed small-scale sunspots and active-region filaments, which also decayed during the disk passage. NOAA~11835 developed fully by day~4 of the disk passage and decayed over the course of the disk passage. However, Fig.~\ref{Fig:mean_int_ar_all} contains the values for the whole disk passage and it is centered on the passage of the active region through the central meridian. In Fig.~\ref{Fig:ar_overview}, we illustrate the isolated patches of H$\alpha$ excess in the mask inside the field-of-view, which were used to calculate the imaging H$\alpha$ excess and its mean intensity. The total number of pixels was derived from all pixels belonging to the mask. The evolution of the imaging H$\alpha$ excess (black curve), the mean intensity of each H$\alpha$ excess region (red curve), and the number of pixels of each region (blue curve) during the disk passage of 15 active regions is displayed in Fig.~\ref{Fig:mean_int_ar_all}. The days are normalized to the passage of the active region through the central meridian. The active regions are roughly organized by their size with the smallest active regions in the top row and the largest active regions in the bottom row. As defined in Eq.~(\ref{eq:excess}), the imaging H$\alpha$ excess is a function of the number of pixels in each mask and the intensity.

For most of the selected active regions, the shape of the number of pixels and the H$\alpha$ excess can be described with a bell shape. The mean intensity varies from the bell-shape structure, but in many cases it is still visible (Fig.~\ref{Fig:mean_int_ar_all}, f, h, i, j, and m). In these cases, the Spearman correlation is also very high between the H$\alpha$ excess and the mean intensity. The correlation between the H$\alpha$ excess and the number of pixels is high in most cases. In ten of the 15 cases, this correlation is the highest of the three calculated correlations. Especially for the larger active regions, the correlation is between $\rho_\mathrm{E, P} = 0.71$ and $\rho_\mathrm{E, P} = 0.97$ (Fig.~\ref{Fig:mean_int_ar_all}, h\,--\,m), which indicates a large dependence of the H$\alpha$ excess on the size of the region. Nonetheless, especially during large flare activity, the mean intensity is largely effected and also influences the H$\alpha$ excess, as is visible for the active regions NOAA~11476 and NOAA~11818 (Fig.~\ref{Fig:mean_int_ar_all}, k and l), where several flares including M-class flares appeared during the disk passage. For NOAA~11818, the Spearman correlation of the H$\alpha$ excess and the mean intensity exceed the correlation of the H$\alpha$ excess with the number of pixels. For active region NOAA~12767 (Fig.~\ref{Fig:mean_int_ar_all}, d), the shape of the three parameters are very similar, but the correlation of the H$\alpha$ excess with the mean intensity is relatively low with  $\rho_\mathrm{E, I} = 0.32$. Therefore, we shifted the mean intensity by one day and repeated the calculation of the correlation, which resulted in a much higher value of $\rho_\mathrm{E, I} = 0.84$. The correlation of the number of pixels with the mean intensity increased to  $\rho_\mathrm{P, I} = 0.75$ as well.  A similar behavior is found for NOAA~11476,  NOAA~12712, and  NOAA~12578 (Fig.~\ref{Fig:mean_int_ar_all}, k, i, and b, respectively).

In addition, we created scatter plots (not shown) of all active regions for each parameter and calculated the Spearman correlation for them. This results in a very high correlation for the H$\alpha$ excess and the number of pixles with $\rho_\mathrm{E, P} = 0.92$. The correlation of the H$\alpha$ excess with the mean intensity is still high, but much lower with $\rho_\mathrm{E, I} = 0.75$. The correlation of the number of pixels and the mean intensity is $\rho_\mathrm{P, I} = 0.45$. By shifting the mean intensity by one day backward, the correlation of the mean intensity and the number of pixels increases to $\rho_\mathrm{P, I} = 0.53$.

%-------------------------------------- One column figure
%-----------------------------------------------------------------

\section{Discussion} \label{sec:disc}

\subsection{A new tracer of the solar activity: Imaging H\texorpdfstring{$\alpha$}{-alpha} excess}  \label{sec:disc_E100}

The prominent H$\alpha$ spectral line is commonly known to reflect solar activity, but it is rarely used as such. The work by \citet{Livingston2007} is one of the very few studies that analyzed the evolution of the H$\alpha$ line-core intensity for Solar Cycle~22 and 23 in comparison to many other spectral indices. The study of \citet{Meunier2009} analyzed the influence of dark absorption features on the integrated H$\alpha$ intensity with respect to its correlation to the \mbox{Ca\,\textsc{ii}\,K} index. In this study, we explore the possibilities of using H$\alpha$ full-disk filtergrams to create a tracer for the solar activity, that is to say the imaging H$\alpha$ excess and deficit.

We compared the positive and negative H$\alpha$ excess with the following already established solar indices: relative sunspot number, F10.7\,cm radio flux, and \mbox{Mg\,\textsc{ii}} index. The highest correlation with the H-alpha excess was found with the \mbox{Mg\,\textsc{ii}} index, which originates in the chromosphere, similar to the H$\alpha$ spectral line. The correlation of the H$\alpha$ deficit and the \mbox{Mg\,\textsc{ii}} index is also very high, but the correlation to the F10.7\,cm radio flux is slightly higher. We also investigated the mean intensity of H$\alpha$ excess regions, which showed that the mean intensity of H$\alpha$ excess regions also changes with the solar cycle. The threshold for the H$\alpha$ excess was selected to include mainly plage regions. The disadvantage of the mean intensity is that the daily variations are very high and the decrease toward the solar minimum is shallow. This is different for the imaging H$\alpha$ excess. Here, the advantage is that it scales with the number of pixels. This results in a steeper decrease in the curve toward the minimum and a better representation of the maximum peaks. In addition, it allows for a better comparison with other activity tracers. The imaging H$\alpha$ excess of bright regions displays the strongest variations in amplitude because bright regions disappear nearly completely from the solar disk during solar minimum. In the master thesis of \citet{LePhuong2016MSC}, an H$\alpha$ spectral index from spectra of the Integrated Sunlight Spectrometer \citep[ISS, ][]{Keller2003} of SOLIS was derived based on the intensity enhancement in the line core. However, the variations during the solar cycle were weak, as indicated in the study of \citet{Maldonado2019}. In any case, solar activity is small compared to stellar activity, which  might be a limiting factor for spectral indices, while disk-resolved indices may perform better.

Not only does the H$\alpha$ excess reflect solar activity, but also the  imaging  H$\alpha$ deficit, which had never been studied before. The trend of the H$\alpha$ deficit, as well as the high correlation to the H$\alpha$ excess and other tracers of solar activity indicate a dependency on the solar cycle. The daily variation of the H$\alpha$ deficit is much stronger in the maximum, covering the full range from maximum values to low values of about $D_{100} \approx 1$ in just a few days, while variations of the daily H$\alpha$ deficit are very small during the minimum. This indicates that absorption features such as filaments are reduced in number during the solar minimum. Dark features are associated with the solar cycle, such as sunspots and filaments. Moreover, polar crown filaments -- large-scale filaments in polar regions -- start to appear in a large number around and after the minimum of the solar cycle until they reach a peak shortly before the magnetic field reversal, that is around solar maximum, when they disappear from the solar disk \citep{Cliver2014, Xu2018, Diercke2019b}. In Fig.~\ref{Fig:index_comp} (upper middle panel), we find that just before the magnetic field reversal in the northern (November 2012) and southern hemispheres (March 2014), the H$\alpha$ deficit is high and drops after the reversal, which indicates the disappearance of polar crown filaments after the magnetic field reversal at the end of the rush-to-the-pole of polar crown filaments. This shows a very good representation of the polar crown cycle with the imaging H$\alpha$ deficit. Nonetheless, the number of absorption structures increases to an even higher level after the magnetic field reversal, leading to the peak of the H$\alpha$ deficit at the end of 2015 (Fig.~\ref{Fig:index_comp}). In the ChroTel H$\alpha$ filtergrams, it is visible that the number of large-scale quiet Sun filaments at mid-latitudes and sunspots and active region filaments in the activity belt increased, which led to the increase in the H$\alpha$ deficit. A second rush-to-the-pole is not observed \citep{Diercke2019b}.

The H$\alpha$ synoptic charts of filaments from several H$\alpha$ surveys were used by \citet{Makarov2000} and \citet{Makarov2001} to calculate an activity index for the large-scale magnetic field indicating a cyclic behavior for several solar cycles \citep{Ermolli2014}. In contrast to this, we extracted the H$\alpha$ deficit directly from H$\alpha$ full-disk filtergrams and we have shown that it can be used as a tracer for solar activity. The H$\alpha$ deficit monitors the coverage of H$\alpha$ absorption features from small-scale active region filaments and sunspots to large-scale quiet Sun and polar crown filaments. Whereas the active region filaments and sunspots are directly related to the solar cycle and can be correlated to the solar activity, the polar crown filaments are an indirect measure of the solar activity, which should not be neglected in a complete view of the solar activity cycle of the Sun.

A close relationship between excess regions and deficit regions can be assumed by comparing the mean intensity of excess and deficit regions (Fig.~\ref{Fig:mean_compare}). Also the daily variations can be relatively high comparing both quantities, the monthly average indicates a closely related trend, with a high anticorrelation. The mean intensity of the excess regions peaks at about the same values. The peaks are related to the maximum of intensity in the northern and southern hemisphere, and additional peaks occur shortly after the maximum in 2015 and another intensity increase in the end of 2017, which is reflected in all solar activity tracers. The peaks of the mean intensity of H$\alpha$ deficit are related to the polar crown activity cycle and after the magnetic field reversal on both hemispheres to an increased number of quiet-Sun and active region filaments.

The distribution of H$\alpha$ excess intensities on the solar disk can be linked to our knowledge of other stars that display activity cycles. Stellar activity cycles are studied in various observables, that is from chromospheric emission \citep{Noyes1984, Baliunas1995, Hall2004, Metcalfe2010} over white light modulation \citep{Reinhold2017, Nielsen2019} to X-ray emission \citep{Hempelmann1996, Sanz-Forcada2013}. Here, we focus on a particularly useful comparison to findings about stellar coronae during stellar activity minima and maxima. While our solar observations refer to the chromosphere, high and low activity regions in the corona are colocated with their counterparts in the chromosphere. We can therefore make some qualitative comparisons to findings from stellar coronae.

Coronal changes over a stellar activity cycle can be observed in soft X-rays, and a small number of cool stars have been monitored over more than a decade for their coronal cycles \citep{Hempelmann2006, Favata2008, Robrade2012, Orlando2017}. When analyzing soft X-ray spectra with moderate resolution, such as they are provided by the current large X-ray observatories XMM-Newton \citep{Jansen2001} and Chandra \citep{Weisskopf2002}, the data quality typically allows for two to three temperature components of the stellar corona to be fitted simultaneously. \citet{Robrade2012} show that for the two stars 61~Cyg~A and $\alpha$~Cen~B, the coronal spectra in both the high and low activity states can be fitted with three temperature components each. When letting the temperatures of the three components free to fit, they found that the fitted temperatures are virtually the same, and only the emission measure of each component changes over the course of the cycle. This is in line with our finding that the distribution of mean intensities of individual H$\alpha$ excess regions retains basically the same shape between solar minimum and maximum, and only the number of regions that are present changes. 

The average solar rotation shows a clear bell-shaped appearance in the maximum phase of the solar cycle, with a higher imaging H$\alpha$ excess and area coverage fraction in the middle of the rotation period compared to the beginning and end (Fig.~\ref{Fig:coverage_all}). For the mean H$\alpha$ excess intensity, the trend is visible but not as pronounced as for the other two parameters. The same trend is seen for the imaging H$\alpha$ deficit and the corresponding area coverage fraction (Fig.~\ref{Fig:coverage_rev_all}). For the declining phase in both cases, this behavior is not noticeable. For single rotation periods, such a trend could be caused by individual large active regions which remain on the solar surface for several solar rotations. Nonetheless, we averaged over 34 Carrington rotations, which should average out such effects. The behavior of sunspots emerging at a preferred longitude is known as active longitude \citep{Berdyugina2003, Hathaway2010, Kramynin2021}. In this study, the active longitude is visible for the H$\alpha$ excess related to plage regions, as well as for the H$\alpha$ deficit, which is mainly related to active region filaments.

In addition, we calculated a strong correlation between the imaging H$\alpha$ excess or deficit with their corresponding area coverage fraction. This indicates that the dominant factor in the calculation of the H$\alpha$ excess and deficit is the number of regions which go into its calculation. The influence of the mean intensity on the H$\alpha$ excess is discussed for individual active regions in the next paragraph.

For most of the active regions selected for this study, the mean intensity (red curve), as well as the number of pixels and imaging H$\alpha$ excess (black curve) show similar temporal profiles, which resemble the bell-shaped profiles of the UV irradiance for active regions presented in \citet{Toriumi2020}, where the 1600\,\AA\ and 1700\,\AA\ images display the transition region and (upper) photosphere. The UV intensity profiles mainly represent signals from bright plage regions, which is also the case for the H$\alpha$ excess. 

The H$\alpha$ excess is strongly correlated to the number of pixels in this region in most cases. Nonetheless, the intensity can influence the H$\alpha$ excess as well, that is during M-class flares, when the intensity is locally enhanced. The correlation of the H$\alpha$ excess and the mean intensity is high in the majority of the active regions. On the other hand, the number of pixels and the mean H$\alpha$ excess intensity do not seem to correlate well. Only in a few cases is the correlation above 0.5.

\subsection{Activity contamination in exoplanet transmission spectra} \label{sec:exo}

The Sun has frequently been used in case studies to understand how exoplanet properties can be affected by the surface features of their host stars \citep[e.g.,][]{meunier10,llama15,haywood16}. Detections of molecules and atoms in exoplanet atmospheres can be contaminated by stellar active regions, which has been coined the "transit light source effect" or "contrast effect" by various authors \citep{cauley18,rackham18}. Indeed, transits of active latitudes can entirely mimic an absorption signal in the planet's atmosphere for strong chromospheric lines such as H$\alpha$ or \ion{He}{1} 10830 \AA\ \citep{cauley17,salz2018,cauley18}. While broadband variability and/or multicomponent spectral fitting can place strong limits on the star spot coverage fraction \citep{gully17} and, in some cases, bright faculae and plage coverage fractions \citep{morris17}, these techniques provide little information concerning the emission line flux from individual bright active regions. Being able to constrain the strength of active region emission for individual stars will provide us with a much more precise contamination estimate for exoplanet atmospheric absorption, that is, how much of the observed signal is due to the occultation of active regions by the planetary disk.

Active region contamination of transmission spectra is only a problem for planets transiting active stars. One group of systems for which this is particularly important is short-period planets transiting very young stars, for example, V1298 Tau \citep{david19}, AU Mic \citep{plavchan20}, and DS Tuc \citep{newton19}. Ultraviolet and H$\alpha$ transmission spectra for these targets can be used to search for signatures of extended atmospheres and evaporation, which requires an understanding of the degree to which stellar active regions affect the planetary signal. A relationship between the area coverage fraction of  magnetic bright regions and the average intensity of the emission from those active regions could be included in models of the rotational modulation of these young stars in order to provide constraints on the two model parameters. This method is especially applicable to young stars given their relatively short rotation periods, which makes it possible to collect data across many rotations. Although a functional form for the relationship between the emission line strength and active region coverage derived for the Sun may not be directly applicable to premain sequence stars or different spectral types.

The H$\alpha$ data presented here offer some insight into this problem. Figure~\ref{Fig:coverage_all} demonstrates that the area coverage fraction of H$\alpha$ excess regions strongly correlates with the imaging H$\alpha$ excess, but the correlation to the mean intensity of H$\alpha$ excess regions is relatively low. For five out of the 15 selected active regions, the correlation is high and for three additional cases the correlation significantly increases when the mean intensity is shifted by one day backward, which would imply a delayed increase in the intensity in these cases. Nonetheless, the anticipated direct relationship of the size of the H$\alpha$ excess regions and their mean intensity cannot be found in the data.

The data were limb darkening corrected and normalized to the median intensity, but they were not corrected for geometrical effects at the limb. Nonetheless, we compared the H$\alpha$ excess, the area coverage fraction, and the mean intensity of H$\alpha$ excess regions between them, so they would be effected in a similar manner by the geometrical distortion. The relationship between them did not change. The same is applicable for the internal evolution of the active region during its disk passage.

\section{Conclusions}

Solar full-disk H$\alpha$ observations are available, almost continuously, from a network of ground-based observatories. They are an ideal source for tracking solar activity. Moreover, H$\alpha$ is commonly used in stellar observations and a comparison of solar and stellar relations is possible. For the solar case, H$\alpha$ is not an established tracer for solar activity yet, but this study should change this by showing some easy quantities, which can be derived from H$\alpha$ full-disk observations.

This study shows the value of the H$\alpha$ spectral window in deriving a tracer for solar activity. The imaging H$\alpha$ excess created from H$\alpha$ excess regions in full-disk filtergrams is closely related to other chromospheric tracers such as the \mbox{Mg\,\textsc{ii}} index. The H$\alpha$ deficit reflects the solar activity throughout the cycle, as well, with a high correlation to the F10.7cm radio flux. The main advantage of the imaging H$\alpha$ excess compared to the mean intensity is a better reflection of the solar activity -- especially compared to other tracers -- which is derived from the strong dependency on the area coverage fraction of H$\alpha$ excess regions. Furthermore, we show that there is a strong correlation of the mean intensity of the H$\alpha$ excess and deficit throughout the cycle, when comparing the monthly average of both quantities. 

We have analyzed the influence of the area coverage fraction and the mean intensity on the imaging H$\alpha$ excess and deficit in detail throughout the rotational period and for a sample of 15 active regions. The correlation with the area coverage fraction was in all cases very high, whereas the correlation to the mean intensity was lower or very small. In a few cases, the correlation was high. A direct relationship between the mean intensity of the H$\alpha$ excess regions and the area coverage fraction was not found. This would have an impact on the modeling of stellar active regions, where the area coverage fraction and the intensity of H$\alpha$ emitting regions are required to accurately represent chromospheres of solar-like stars. Furthermore, the active latitude of sunspots is visible with the H$\alpha$ excess and deficit  in a Carrington reference frame averaged over the maximum phase of the solar cycle. The histogram of the mean intensity of H$\alpha$ excess regions for the maximum and minimum of the solar cycle revealed a similar shape, but it was reduced by a factor of four in the minimum of the solar cycle.

\begin{acknowledgements}
ChroTel is operated by the Leibniz-Institut f\"ur Sonnenphysik (KIS) in Freiburg, Germany, at the Spanish Observatorio del Teide on Tenerife (Spain). The ChroTel filtergraph was developed by the KIS in cooperation with the High Altitude Observatory (HAO) in Boulder, Colorado. The National Solar Observatory (NSO) is operated by the Association of Universities for Research in Astronomy, Inc. (AURA), under cooperative agreement with the National Science Foundation.  CD acknowledges the support by grant DE~787/5-1  of the  \textit{Deutsche Forschungsgemeinschaft} (DFG). CD and CK acknowledge the support by the  \textit{European Commission's Horizon 2020 Program} under grant agreements 824064 (ESCAPE -- European Science Cluster of Astronomy \& Particle physics ESFRI research infrastructures) and 824135 (SOLARNET -- Integrating High Resolution Solar Physics). KP acknowledges support from the \textit{Leibniz-Gemeinschaft} under grant P67/2018. ED is grateful for financial support from the \textit{Deutscher Akademischer Austauschdienst} (DAAD) enabling her participation in the \textit{Leibniz Graduate School for Quantitative Spectroscopy in Astrophysics}. The authors thank Dr. A. Tritschler for the helpful comments on the manuscript. This research has made use of NASA's Astrophysics Data System. We thank the anonymous referee for the valuable comments significantly improving the manuscript.
\end{acknowledgements}

% WARNING
%-------------------------------------------------------------------
% Please note that we have included the references to the file aa.dem in
% order to compile it, but we ask you to:
%
% - use BibTeX with the regular commands:
%\bibliographystyle{aa} % style aa.bst
%\bibliography{aa-jour,halpha} % your references Yourfile.bib

\begin{thebibliography}{92}
\expandafter\ifx\csname natexlab\endcsname\relax\def\natexlab#1{#1}\fi

\bibitem[{{Arlt} {et~al.}(2013){Arlt}, {Leussu}, {Giese}, {Mursula}, \&
  {Usoskin}}]{Arlt2013}
{Arlt}, R., {Leussu}, R., {Giese}, N., {Mursula}, K., \& {Usoskin}, I.~G. 2013,
  MNRAS, 433, 3165

\bibitem[{{Arlt} \& {Vaquero}(2020)}]{Arlt2020}
{Arlt}, R. \& {Vaquero}, J.~M. 2020, Living Rev. Sol. Phys., 17, 1

\bibitem[{{Ayres}(1989)}]{Ayres1989}
{Ayres}, T.~R. 1989, Sol.\ Phys., 124, 15

\bibitem[{{Baliunas} {et~al.}(1995){Baliunas}, {Donahue}, {Soon}, {Horne},
  {Frazer}, {Woodard-Eklund}, {Bradford}, {Rao}, {Wilson}, {Zhang}, {Bennett},
  {Briggs}, {Carroll}, {Duncan}, {Figueroa}, {Lanning}, {Misch}, {Mueller},
  {Noyes}, {Poppe}, {Porter}, {Robinson}, {Russell}, {Shelton}, {Soyumer},
  {Vaughan}, \& {Whitney}}]{Baliunas1995}
{Baliunas}, S.~L., {Donahue}, R.~A., {Soon}, W.~H., {et~al.} 1995, \apj, 438,
  269

\bibitem[{{Barczynski} {et~al.}(2018){Barczynski}, {Peter}, {Chitta}, \&
  {Solanki}}]{Barczynski2018}
{Barczynski}, K., {Peter}, H., {Chitta}, L.~P., \& {Solanki}, S.~K. 2018, A\&A,
  619, A5

\bibitem[{{Berdyugina} \& {Usoskin}(2003)}]{Berdyugina2003}
{Berdyugina}, S.~V. \& {Usoskin}, I.~G. 2003, A\&A, 405, 1121

\bibitem[{{Bertello} {et~al.}(2016){Bertello}, {Pevtsov}, {Tlatov}, \&
  {Singh}}]{Bertello2016}
{Bertello}, L., {Pevtsov}, A., {Tlatov}, A., \& {Singh}, J. 2016, Sol.\ Phys.,
  291, 2967

\bibitem[{{Bethge} {et~al.}(2011){Bethge}, {Peter}, {Kentischer},
  {Halbgewachs}, {Elmore}, \& {Beck}}]{Bethge2011}
{Bethge}, C., {Peter}, H., {Kentischer}, T.~J., {et~al.} 2011, A\&A, 534, A105

\bibitem[{{Carrington}(1858)}]{Carrington1858}
{Carrington}, R.~C. 1858, MNRAS, 19, 1

\bibitem[{{Cauley} {et~al.}(2018){Cauley}, {Kuckein}, {Redfield}, {Shkolnik},
  {Denker}, {Llama}, \& {Verma}}]{cauley18}
{Cauley}, P.~W., {Kuckein}, C., {Redfield}, S., {et~al.} 2018, \aj, 156, 189

\bibitem[{{Cauley} {et~al.}(2017){Cauley}, {Redfield}, \& {Jensen}}]{cauley17}
{Cauley}, P.~W., {Redfield}, S., \& {Jensen}, A.~G. 2017, \aj, 153, 217

\bibitem[{{Chatzistergos} {et~al.}(2020){Chatzistergos}, {Ermolli}, {Krivova},
  {Solanki}, {Banerjee}, {Barata}, {Belik}, {Gafeira}, {Garcia}, {Hanaoka},
  {Hegde}, {Klime{\v{s}}}, {Korokhin}, {Louren{\c{c}}o}, {Malherbe},
  {Marchenko}, {Peixinho}, {Sakurai}, \& {Tlatov}}]{Chatzistergos2020a}
{Chatzistergos}, T., {Ermolli}, I., {Krivova}, N.~A., {et~al.} 2020, A\&A, 639,
  A88

\bibitem[{{Cliver}(2014)}]{Cliver2014}
{Cliver}, E.~W. 2014, SSR, 186, 169

\bibitem[{{David} {et~al.}(2019){David}, {Petigura}, {Luger}, {Foreman-Mackey},
  {Livingston}, {Mamajek}, \& {Hillenbrand}}]{david19}
{David}, T.~J., {Petigura}, E.~A., {Luger}, R., {et~al.} 2019, \apjl, 885, L12

\bibitem[{{Deng} {et~al.}(2015){Deng}, {Zhang}, {Wang}, {Ji}, {Wang}, {Liu},
  {Xiang}, {Jin}, \& {Cao}}]{Deng2015}
{Deng}, H., {Zhang}, D., {Wang}, T., {et~al.} 2015, Sol.\ Phys., 290, 1479

\bibitem[{{Denker} {et~al.}(2018){Denker}, {Dineva}, {Balthasar}, {Verma},
  {Kuckein}, {Diercke}, \& {Gonz{\'a}lez Manrique}}]{Denker2018}
{Denker}, C., {Dineva}, E., {Balthasar}, H., {et~al.} 2018, Sol.\ Phys., 5, 236

\bibitem[{{Denker} {et~al.}(1999){Denker}, {Johannesson}, {Marquette}, {Goode},
  {Wang}, \& {Zirin}}]{Denker1999}
{Denker}, C., {Johannesson}, A., {Marquette}, W., {et~al.} 1999, Sol.\ Phys.,
  184, 87

\bibitem[{{Denker} \& {Tritschler}(2005)}]{Denker2005}
{Denker}, C. \& {Tritschler}, A. 2005, Publ. Astron. Soc. Pac., 117, 1435

\bibitem[{{Diercke} {et~al.}(2015){Diercke}, {Arlt}, \& {Denker}}]{Diercke2015}
{Diercke}, A., {Arlt}, R., \& {Denker}, C. 2015, AN, 336, 53

\bibitem[{{Diercke} \& {Denker}(2019)}]{Diercke2019b}
{Diercke}, A. \& {Denker}, C. 2019, Sol.\ Phys., 294, 152

\bibitem[{{Diercke} {et~al.}(2018){Diercke}, {Kuckein}, {Verma}, \&
  {Denker}}]{Diercke2018}
{Diercke}, A., {Kuckein}, C., {Verma}, M., \& {Denker}, C. 2018, A\&A, 611, A64

\bibitem[{{Ermolli} {et~al.}(2014){Ermolli}, {Shibasaki}, {Tlatov}, \& {van
  Driel-Gesztelyi}}]{Ermolli2014}
{Ermolli}, I., {Shibasaki}, K., {Tlatov}, A., \& {van Driel-Gesztelyi}, L.
  2014, SSR, 186, 105

\bibitem[{{Ermolli} {et~al.}(2015){Ermolli}, {Shibasaki}, {Tlatov}, \& {van
  Driel-Gesztelyi}}]{Ermolli2015}
{Ermolli}, I., {Shibasaki}, K., {Tlatov}, A., \& {van Driel-Gesztelyi}, L.
  2015, in The Solar Activity Cycle, ed. A.~{Balogh}, H.~{Hudson},
  K.~{Petrovay}, \& R.~{von Steiger}, Vol.~53 (New York, NY, USA: Springer),
  105--135

\bibitem[{{Favata} {et~al.}(2008){Favata}, {Micela}, {Orlando}, {Schmitt},
  {Sciortino}, \& {Hall}}]{Favata2008}
{Favata}, F., {Micela}, G., {Orlando}, S., {et~al.} 2008, \aap, 490, 1121

\bibitem[{{Freytag} {et~al.}(2002){Freytag}, {Steffen}, \&
  {Dorch}}]{Freytag2002}
{Freytag}, B., {Steffen}, M., \& {Dorch}, B. 2002, AN, 323, 213

\bibitem[{{Fr{\"o}hlich}(2012)}]{Froehlich2012}
{Fr{\"o}hlich}, C. 2012, Surveys in Geophysics, 33, 453

\bibitem[{{Golub} {et~al.}(2007){Golub}, {Deluca}, {Austin}, {Bookbinder},
  {Caldwell}, {Cheimets}, {Cirtain}, {Cosmo}, {Reid}, {Sette}, {Weber},
  {Sakao}, {Kano}, {Shibasaki}, {Hara}, {Tsuneta}, {Kumagai}, {Tamura},
  {Shimojo}, {McCracken}, {Carpenter}, {Haight}, {Siler}, {Wright}, {Tucker},
  {Rutledge}, {Barbera}, {Peres}, \& {Varisco}}]{Golub2007}
{Golub}, L., {Deluca}, E., {Austin}, G., {et~al.} 2007, Sol.\ Phys., 243, 63

\bibitem[{{Gully-Santiago} {et~al.}(2017){Gully-Santiago}, {Herczeg},
  {Czekala}, {Somers}, {Grankin}, {Covey}, {Donati}, {Alencar}, {Hussain},
  {Shappee}, {Mace}, {Lee}, {Holoien}, {Jose}, \& {Liu}}]{gully17}
{Gully-Santiago}, M.~A., {Herczeg}, G.~J., {Czekala}, I., {et~al.} 2017, \apj,
  836, 200

\bibitem[{{Haberreiter} {et~al.}(2021){Haberreiter}, {Criscuoli}, {Rempel}, \&
  {Pereira}}]{Haberreiter2021}
{Haberreiter}, M., {Criscuoli}, S., {Rempel}, M., \& {Pereira}, T. M.~D. 2021,
  A\&A, 653, A161

\bibitem[{{Hall} \& {Lockwood}(2004)}]{Hall2004}
{Hall}, J.~C. \& {Lockwood}, G.~W. 2004, \apj, 614, 942

\bibitem[{{Hanaoka} \& {Sakurai}(2020)}]{Hanaoka2020}
{Hanaoka}, Y. \& {Sakurai}, T. 2020, ApJ, 904, 63

\bibitem[{{Harvey} {et~al.}(1996){Harvey}, {Hill}, {Hubbard}, {Kennedy},
  {Leibacher}, {Pintar}, {Gilman}, {Noyes}, {Title}, {Toomre}, {Ulrich},
  {Bhatnagar}, {Kennewell}, {Marquette}, {Patron}, {Saa}, \&
  {Yasukawa}}]{Harvey1996}
{Harvey}, J.~W., {Hill}, F., {Hubbard}, R.~P., {et~al.} 1996, Sci, 272, 1284

\bibitem[{{Hathaway}(2010)}]{Hathaway2010}
{Hathaway}, D.~H. 2010, Liv. Rev. Sol. Phys., 7, 1

\bibitem[{{Haywood} {et~al.}(2016){Haywood}, {Collier Cameron}, {Unruh},
  {Lovis}, {Lanza}, {Llama}, {Deleuil}, {Fares}, {Gillon}, {Moutou}, {Pepe},
  {Pollacco}, {Queloz}, \& {S{\'e}gransan}}]{haywood16}
{Haywood}, R.~D., {Collier Cameron}, A., {Unruh}, Y.~C., {et~al.} 2016, \mnras,
  457, 3637

\bibitem[{{Hempelmann} {et~al.}(2006){Hempelmann}, {Robrade}, {Schmitt},
  {Favata}, {Baliunas}, \& {Hall}}]{Hempelmann2006}
{Hempelmann}, A., {Robrade}, J., {Schmitt}, J.~H.~M.~M., {et~al.} 2006, \aap,
  460, 261

\bibitem[{{Hempelmann} {et~al.}(1996){Hempelmann}, {Schmitt}, \&
  {St{\c{e}}pie{\'n}}}]{Hempelmann1996}
{Hempelmann}, A., {Schmitt}, J.~H.~M.~M., \& {St{\c{e}}pie{\'n}}, K. 1996,
  \aap, 305, 284

\bibitem[{{Holzreuter} \& {Solanki}(2015)}]{Holzreuter15}
{Holzreuter}, R. \& {Solanki}, S.~K. 2015, \aap, 582, A101

\bibitem[{{Jansen} {et~al.}(2001){Jansen}, {Lumb}, {Altieri}, {Clavel}, {Ehle},
  {Erd}, {Gabriel}, {Guainazzi}, {Gondoin}, {Much}, {Munoz}, {Santos},
  {Schartel}, {Texier}, \& {Vacanti}}]{Jansen2001}
{Jansen}, F., {Lumb}, D., {Altieri}, B., {et~al.} 2001, \aap, 365, L1

\bibitem[{{Johannesson} {et~al.}(1998){Johannesson}, {Marquette}, \&
  {Zirin}}]{Johannesson1998}
{Johannesson}, A., {Marquette}, W.~H., \& {Zirin}, H. 1998, Sol. Phys., 177,
  265

\bibitem[{{Keller} {et~al.}(2003){Keller}, {Harvey}, \&
  {Giampapa}}]{Keller2003}
{Keller}, C.~U., {Harvey}, J.~W., \& {Giampapa}, M.~S. 2003, in Proc. SPIE,
  Vol. 4853, Innovative Telescopes and Instrumentation for Solar Astrophysics,
  ed. S.~L. {Keil} \& S.~V. {Avakyan}, 194--204

\bibitem[{{Kentischer} {et~al.}(2008){Kentischer}, {Bethge}, {Elmore},
  {Friedlein}, {Halbgewachs}, {Kn{\"o}lker}, {Peter}, {Schmidt}, {Sigwarth}, \&
  {Streander}}]{Kentischer2008}
{Kentischer}, T.~J., {Bethge}, C., {Elmore}, D.~F., {et~al.} 2008, in Proc.
  SPIE, Vol. 7014, Ground-based and Airborne Instrumentation for Astronomy II,
  ed. I.~S. {McLean} \& M.~M. {Casali}, 701413

\bibitem[{{Knaack} {et~al.}(2004){Knaack}, {Stenflo}, \&
  {Berdyugina}}]{Knaack2004}
{Knaack}, R., {Stenflo}, J.~O., \& {Berdyugina}, S.~V. 2004, A\&A, 418, L17

\bibitem[{{Kramynin} \& {Mikhalina}(2021)}]{Kramynin2021}
{Kramynin}, A.~P. \& {Mikhalina}, F.~A. 2021, Geomagnetism and Aeronomy, 61,
  937

\bibitem[{{Kuckein} {et~al.}(2016){Kuckein}, {Verma}, \&
  {Denker}}]{Kuckein2016}
{Kuckein}, C., {Verma}, M., \& {Denker}, C. 2016, A\&A, 589, A84

\bibitem[{{Le Phuong}(2016)}]{LePhuong2016MSC}
{Le Phuong}, L. 2016, {Master Thesis}, Universit\"at Potsdam, Germany

\bibitem[{{Lean} {et~al.}(1997){Lean}, {Rottman}, {Kyle}, {Woods}, {Hickey}, \&
  {Puga}}]{Lean1997}
{Lean}, J.~L., {Rottman}, G.~J., {Kyle}, H.~L., {et~al.} 1997, J. Geophys.
  Res., 102, 29939

\bibitem[{{Leenaarts} {et~al.}(2012){Leenaarts}, {Carlsson}, \& {Rouppe van der
  Voort}}]{leenaarts12}
{Leenaarts}, J., {Carlsson}, M., \& {Rouppe van der Voort}, L. 2012, \apj, 749,
  136

\bibitem[{{Lemen} {et~al.}(2012){Lemen}, {Title}, {Akin}, {Chou}, {Drake},
  {Duncan}, {Edwards}, {Friedlaender}, {Heyman}, {Hurlburt}, {Katz}, {Kushner},
  {Levay}, {Lindgren}, {Mathur}, {McFeaters}, {Mitchell}, {Rehse}, {Schrijver},
  {Springer}, {Stern}, {Tarbell}, {Wuelser}, {Wolfson}, {Yanari}, {Bookbinder},
  {Cheimets}, {Caldwell}, {Deluca}, {Gates}, {Golub}, {Park}, {Podgorski},
  {Bush}, {Scherrer}, {Gummin}, {Smith}, {Auker}, {Jerram}, {Pool}, {Soufli},
  {Windt}, {Beardsley}, {Clapp}, {Lang}, \& {Waltham}}]{Lemen2012}
{Lemen}, J.~R., {Title}, A.~M., {Akin}, D. J.and~{Boerner}, P.~F., {et~al.}
  2012, Sol.\ Phys., 275, 17

\bibitem[{{Leroy} {et~al.}(1983){Leroy}, {Bommier}, \&
  {Sahal-Brechot}}]{Leroy1983}
{Leroy}, J.~L., {Bommier}, V., \& {Sahal-Brechot}, S. 1983, Sol.\ Phys., 83,
  135

\bibitem[{{Leroy} {et~al.}(1984){Leroy}, {Bommier}, \&
  {Sahal-Brechot}}]{Leroy1984}
{Leroy}, J.~L., {Bommier}, V., \& {Sahal-Brechot}, S. 1984, A\&A, 131, 33

\bibitem[{{Livingston} {et~al.}(2007){Livingston}, {Wallace}, {White}, \&
  {Giampapa}}]{Livingston2007}
{Livingston}, W., {Wallace}, L., {White}, O.~R., \& {Giampapa}, M.~S. 2007,
  ApJ, 657, 1137

\bibitem[{{Llama} \& {Shkolnik}(2015)}]{llama15}
{Llama}, J. \& {Shkolnik}, E.~L. 2015, \apj, 802, 41

\bibitem[{{Makarov} \& {Tlatov}(2000)}]{Makarov2000}
{Makarov}, V.~I. \& {Tlatov}, A.~G. 2000, Astronomy Reports, 44, 759

\bibitem[{{Makarov} {et~al.}(2001){Makarov}, {Tlatov}, {Callebaut}, {Obridko},
  \& {Shelting}}]{Makarov2001}
{Makarov}, V.~I., {Tlatov}, A.~G., {Callebaut}, D.~K., {Obridko}, V.~N., \&
  {Shelting}, B.~D. 2001, Sol.\ Phys., 198, 409

\bibitem[{{Maldonado} {et~al.}(2019){Maldonado}, {Phillips}, {Dumusque},
  {Collier Cameron}, {Haywood}, {Lanza}, {Micela}, {Mortier}, {Saar},
  {Sozzetti}, {Rice}, {Milbourne}, {Cecconi}, {Cegla}, {Cosentino}, {Costes},
  {Ghedina}, {Gonzalez}, {Guerra}, {Hern{\'a}ndez}, {Li}, {Lodi}, {Malavolta},
  {Molinari}, {Pepe}, {Piotto}, {Poretti}, {Sasselov}, {San Juan}, {Thompson},
  {Udry}, \& {Watson}}]{Maldonado2019}
{Maldonado}, J., {Phillips}, D.~F., {Dumusque}, X., {et~al.} 2019, A\&A, 627,
  A118

\bibitem[{{Metcalfe} {et~al.}(2010){Metcalfe}, {Basu}, {Henry}, {Soderblom},
  {Judge}, {Kn{\"o}lker}, {Mathur}, \& {Rempel}}]{Metcalfe2010}
{Metcalfe}, T.~S., {Basu}, S., {Henry}, T.~J., {et~al.} 2010, \apjl, 723, L213

\bibitem[{{Meunier} \& {Delfosse}(2009)}]{Meunier2009}
{Meunier}, N. \& {Delfosse}, X. 2009, A\&A, 501, 1103

\bibitem[{{Meunier} {et~al.}(2010){Meunier}, {Desort}, \&
  {Lagrange}}]{meunier10}
{Meunier}, N., {Desort}, M., \& {Lagrange}, A.~M. 2010, \aap, 512, A39

\bibitem[{{Morris} {et~al.}(2017){Morris}, {Hebb}, {Davenport}, {Rohn}, \&
  {Hawley}}]{morris17}
{Morris}, B.~M., {Hebb}, L., {Davenport}, J. R.~A., {Rohn}, G., \& {Hawley},
  S.~L. 2017, \apj, 846, 99

\bibitem[{{Naqvi} {et~al.}(2010){Naqvi}, {Marquette}, {Tritschler}, \&
  {Denker}}]{Naqvi2010}
{Naqvi}, M.~F., {Marquette}, W.~H., {Tritschler}, A., \& {Denker}, C. 2010, AN,
  331, 696

\bibitem[{{Newton} {et~al.}(2019){Newton}, {Mann}, {Tofflemire}, {Pearce},
  {Rizzuto}, {Vanderburg}, {Martinez}, {Wang}, {Ruffio}, {Kraus}, {Johnson},
  {Thao}, {Wood}, {Rampalli}, {Nielsen}, {Collins}, {Dragomir}, {Hellier},
  {Anderson}, {Barclay}, {Brown}, {Feiden}, {Hart}, {Isopi}, {Kielkopf},
  {Mallia}, {Nelson}, {Rodriguez}, {Stockdale}, {Waite}, {Wright}, {Lissauer},
  {Ricker}, {Vanderspek}, {Latham}, {Seager}, {Winn}, {Jenkins}, {Bouma},
  {Burke}, {Davies}, {Fausnaugh}, {Li}, {Morris}, {Mukai}, {Villase{\~n}or},
  {Villeneuva}, {De Rosa}, {Macintosh}, {Mengel}, {Okumura}, \&
  {Wittenmyer}}]{newton19}
{Newton}, E.~R., {Mann}, A.~W., {Tofflemire}, B.~M., {et~al.} 2019, \apjl, 880,
  L17

\bibitem[{{Nielsen} {et~al.}(2019){Nielsen}, {Gizon}, {Cameron}, \&
  {Miesch}}]{Nielsen2019}
{Nielsen}, M.~B., {Gizon}, L., {Cameron}, R.~H., \& {Miesch}, M. 2019, \aap,
  622, A85

\bibitem[{{Noyes} {et~al.}(1984){Noyes}, {Hartmann}, {Baliunas}, {Duncan}, \&
  {Vaughan}}]{Noyes1984}
{Noyes}, R.~W., {Hartmann}, L.~W., {Baliunas}, S.~L., {Duncan}, D.~K., \&
  {Vaughan}, A.~H. 1984, \apj, 279, 763

\bibitem[{{Orlando} {et~al.}(2017){Orlando}, {Favata}, {Micela}, {Sciortino},
  {Maggio}, {Schmitt}, {Robrade}, \& {Mittag}}]{Orlando2017}
{Orlando}, S., {Favata}, F., {Micela}, G., {et~al.} 2017, \aap, 605, A19

\bibitem[{{Otruba}(1999)}]{Otruba1999}
{Otruba}, W. 1999, in Third Advances in Solar Physics Euroconference: Magnetic
  Fields and Oscillations, Vol. 184, 314--318

\bibitem[{{Pesnell} {et~al.}(2012){Pesnell}, {Thompson}, \&
  {Chamberlin}}]{Pesnell2012}
{Pesnell}, W.~D., {Thompson}, B.~J., \& {Chamberlin}, P.~C. 2012, Sol.\ Phys.,
  275, 3

\bibitem[{{Plavchan} {et~al.}(2020){Plavchan}, {Barclay}, {Gagn{\'e}}, {Gao},
  {Cale}, {Matzko}, {Dragomir}, {Quinn}, {Feliz}, {Stassun}, {Crossfield},
  {Berardo}, {Latham}, {Tieu}, {Anglada-Escud{\'e}}, {Ricker}, {Vanderspek},
  {Seager}, {Winn}, {Jenkins}, {Rinehart}, {Krishnamurthy}, {Dynes}, {Doty},
  {Adams}, {Afanasev}, {Beichman}, {Bottom}, {Bowler}, {Brinkworth}, {Brown},
  {Cancino}, {Ciardi}, {Clampin}, {Clark}, {Collins}, {Davison},
  {Foreman-Mackey}, {Furlan}, {Gaidos}, {Geneser}, {Giddens}, {Gilbert},
  {Hall}, {Hellier}, {Henry}, {Horner}, {Howard}, {Huang}, {Huber}, {Kane},
  {Kenworthy}, {Kielkopf}, {Kipping}, {Klenke}, {Kruse}, {Latouf}, {Lowrance},
  {Mennesson}, {Mengel}, {Mills}, {Morton}, {Narita}, {Newton}, {Nishimoto},
  {Okumura}, {Palle}, {Pepper}, {Quintana}, {Roberge}, {Roccatagliata},
  {Schlieder}, {Tanner}, {Teske}, {Tinney}, {Vanderburg}, {von Braun}, {Walp},
  {Wang}, {Wang}, {Weigand }, {White}, {Wittenmyer}, {Wright}, {Youngblood},
  {Zhang}, \& {Zilberman}}]{plavchan20}
{Plavchan}, P., {Barclay}, T., {Gagn{\'e}}, J., {et~al.} 2020, \nat, 582, 497

\bibitem[{{Rackham} {et~al.}(2018){Rackham}, {Apai}, \& {Giampapa}}]{rackham18}
{Rackham}, B.~V., {Apai}, D., \& {Giampapa}, M.~S. 2018, \apj, 853, 122

\bibitem[{{Reinhold} {et~al.}(2017){Reinhold}, {Cameron}, \&
  {Gizon}}]{Reinhold2017}
{Reinhold}, T., {Cameron}, R.~H., \& {Gizon}, L. 2017, \aap, 603, A52

\bibitem[{{Rezaei} {et~al.}(2007){Rezaei}, {Schlichenmaier}, {Beck}, {Bruls},
  \& {Schmidt}}]{Rezaei2007}
{Rezaei}, R., {Schlichenmaier}, R., {Beck}, C. A.~R., {Bruls}, J. H. M.~J., \&
  {Schmidt}, W. 2007, A\&A, 466, 1131

\bibitem[{{Robrade} {et~al.}(2012){Robrade}, {Schmitt}, \&
  {Favata}}]{Robrade2012}
{Robrade}, J., {Schmitt}, J.~H.~M.~M., \& {Favata}, F. 2012, \aap, 543, A84

\bibitem[{{Salz} {et~al.}(2018){Salz}, {Czesla}, {Schneider}, {Nagel},
  {Schmitt}, {Nortmann}, {Alonso-Floriano}, {L{\'o}pez-Puertas}, {Lamp{\'o}n},
  {Bauer}, {Snellen}, {Pall{\'e}}, {Caballero}, {Yan}, {Chen}, {Sanz-Forcada},
  {Amado}, {Quirrenbach}, {Ribas}, {Reiners}, {B{\'e}jar}, {Casasayas-Barris},
  {Cort{\'e}s-Contreras}, {Dreizler}, {Guenther}, {Henning}, {Jeffers},
  {Kaminski}, {K{\"u}rster}, {Lafarga}, {Lara}, {Molaverdikhani}, {Montes},
  {Morales}, {S{\'a}nchez-L{\'o}pez}, {Seifert}, {Zapatero Osorio}, \&
  {Zechmeister}}]{salz2018}
{Salz}, M., {Czesla}, S., {Schneider}, P.~C., {et~al.} 2018, \aap, 620, A97

\bibitem[{{Sanz-Forcada} {et~al.}(2013){Sanz-Forcada}, {Stelzer}, \&
  {Metcalfe}}]{Sanz-Forcada2013}
{Sanz-Forcada}, J., {Stelzer}, B., \& {Metcalfe}, T.~S. 2013, \aap, 553, L6

\bibitem[{{Scherrer} {et~al.}(2012){Scherrer}, {Schou}, {Bush}, {Kosovichev},
  {Bogart}, {Hoeksema}, {Liu}, {Duvall}, {Zhao}, {Title}, {Schrijver},
  {Tarbell}, \& {Tomczyk}}]{Scherrer2012}
{Scherrer}, P.~H., {Schou}, J., {Bush}, R.~I., {et~al.} 2012, Sol.\ Phys., 275,
  207

\bibitem[{{Schrijver} {et~al.}(1989){Schrijver}, {Cote}, {Zwaan}, \&
  {Saar}}]{Schrijver1989}
{Schrijver}, C.~J., {Cote}, J., {Zwaan}, C., \& {Saar}, S.~H. 1989, ApJ, 337,
  964

\bibitem[{{Schwabe}(1844)}]{Schwabe1844}
{Schwabe}, H. 1844, AN, 21, 233

\bibitem[{{Sheminova}(2012)}]{Sheminova2012}
{Sheminova}, V.~A. 2012, Sol.\ Phys., 280, 83

\bibitem[{{Shen} {et~al.}(2018){Shen}, {Diercke}, \& {Denker}}]{Shen2018}
{Shen}, Z., {Diercke}, A., \& {Denker}, C. 2018, AN, 339, 661

\bibitem[{{Sim{\~o}es} {et~al.}(2019){Sim{\~o}es}, {Reid}, {Milligan}, \&
  {Fletcher}}]{Simoes2019}
{Sim{\~o}es}, P. J.~A., {Reid}, H. A.~S., {Milligan}, R.~O., \& {Fletcher}, L.
  2019, ApJ, 870, 114

\bibitem[{{Sp{\"o}rer}(1879)}]{Spoerer1879}
{Sp{\"o}rer}, F. W.~G. 1879, AN, 96, 23

\bibitem[{{Steinegger} {et~al.}(2000){Steinegger}, {Denker}, {Goode},
  {Marquette}, {Varsik}, {Wang}, {Otruba}, {Freislich}, {Hanslmeier}, {Luo},
  {Chen}, \& {Zhang}}]{Steinegger2000}
{Steinegger}, M., {Denker}, C., {Goode}, P.~R., {et~al.} 2000, in ESA Special
  Publication, Vol. 463, The Solar Cycle and Terrestrial Climate, Solar and
  Space Weather, ed. A.~{Wilson}, 617

\bibitem[{{Sun} {et~al.}(2015){Sun}, {Hoeksema}, {Liu}, \& {Zhao}}]{Sun2015}
{Sun}, X., {Hoeksema}, J.~T., {Liu}, Y., \& {Zhao}, J. 2015, ApJ, 798, 114

\bibitem[{{Tapping} \& {Charrois}(1994)}]{Tapping1994}
{Tapping}, K.~F. \& {Charrois}, D.~P. 1994, Sol.\ Phys., 150, 305

\bibitem[{{Toriumi} {et~al.}(2020){Toriumi}, {Airapetian}, {Hudson},
  {Schrijver}, {Cheung}, \& {DeRosa}}]{Toriumi2020}
{Toriumi}, S., {Airapetian}, V.~S., {Hudson}, H.~S., {et~al.} 2020, ApJ, 902,
  36

\bibitem[{{van Driel-Gesztelyi} \& {Green}(2015)}]{vanDrielGesztelyi2015}
{van Driel-Gesztelyi}, L. \& {Green}, L.~M. 2015, Living Reviews in Solar
  Physics, 12, 1

\bibitem[{{Verbeeck} {et~al.}(2014){Verbeeck}, {Delouille}, {Mampaey}, \& {De
  Visscher}}]{Verbeeck2014}
{Verbeeck}, C., {Delouille}, V., {Mampaey}, B., \& {De Visscher}, R. 2014,
  A\&A, 561, A29

\bibitem[{{Viereck} \& {Puga}(1999)}]{Viereck1999}
{Viereck}, R.~A. \& {Puga}, L.~C. 1999, J. Geophys. Res., 104, 9995

\bibitem[{{{\v{S}}t{\v{e}}p{\'a}n} {et~al.}(2015){{\v{S}}t{\v{e}}p{\'a}n},
  {Trujillo Bueno}, {Leenaarts}, \& {Carlsson}}]{stepan15}
{{\v{S}}t{\v{e}}p{\'a}n}, J., {Trujillo Bueno}, J., {Leenaarts}, J., \&
  {Carlsson}, M. 2015, \apj, 803, 65

\bibitem[{{Weisskopf} {et~al.}(2002){Weisskopf}, {Brinkman}, {Canizares},
  {Garmire}, {Murray}, \& {Van Speybroeck}}]{Weisskopf2002}
{Weisskopf}, M.~C., {Brinkman}, B., {Canizares}, C., {et~al.} 2002, \pasp, 114,
  1

\bibitem[{{White} \& {Livingston}(1981)}]{White1981}
{White}, O.~R. \& {Livingston}, W.~C. 1981, ApJ, 249, 798

\bibitem[{{Xu} {et~al.}(2021){Xu}, {Banerjee}, {Chatterjee}, {P{\"o}tzi},
  {Wang}, {Ruan}, {Jing}, \& {Wang}}]{Xu2021}
{Xu}, Y., {Banerjee}, D., {Chatterjee}, S., {et~al.} 2021, ApJ, 909, 86

\bibitem[{{Xu} {et~al.}(2018){Xu}, {P{\"o}tzi}, {Zhang}, {Huang}, {Jing}, \&
  {Wang}}]{Xu2018}
{Xu}, Y., {P{\"o}tzi}, W., {Zhang}, H., {et~al.} 2018, ApJ, 862, L23

\end{thebibliography}

%
% - join the .bib files when you upload your source files
%-------------------------------------------------------------------

\end{document}